\documentclass[sigconf]{acmart}
\usepackage{setspace}
\usepackage[utf8]{inputenc}
\usepackage{algorithm}
\usepackage{algorithmic}
\usepackage{amsthm}
\usepackage{float}
\usepackage{graphicx}
\usepackage{booktabs}
\usepackage{comment}
\usepackage{color}  
\usepackage{hyperref}


\usepackage[english]{babel}
\copyrightyear{2021}
\acmYear{2021}
\setcopyright{acmcopyright}
\acmConference[IDEAS 2021]{25th International Database Engineering \& Applications Symposium}{July 14--16, 2021}{Montreal, QC, Canada}
\acmBooktitle{25th International Database Engineering \& Applications Symposium (IDEAS 2021), July 14--16, 2021, Montreal, QC, Canada}
\acmPrice{15.00}
\acmDOI{10.1145/3472163.3472268}
\acmISBN{978-1-4503-8991-4/21/07}

\begin{document}

\title{An Automatic Schema-Instance Approach for Merging Multidimensional Data Warehouses}

\author{Yuzhao Yang}
\affiliation{
\institution{IRIT-CNRS (UMR 5505), 
    Université de Toulouse}
       \city{Toulouse}
       \country{France}
       }
 \email{yuzhao.yang@irit.fr}
 
 \author{Jérôme Darmont}
\affiliation{
\institution{Université de Lyon, Lyon 2, UR ERIC}
       \city{Lyon}
       \country{France}
       }
 \email{jerome.darmont@univ-lyon2.fr}
 
 \author{Franck Ravat}
\affiliation{
\institution{IRIT-CNRS (UMR 5505), 
    Université de Toulouse}
       \city{Toulouse}
       \country{France}
       }
 \email{franck.ravat@irit.fr}
 
 \author{Olivier Teste}
\affiliation{
\institution{IRIT-CNRS (UMR 5505), 
    Université de Toulouse}
       \city{Toulouse}
       \country{France}
       }
 \email{olivier.teste@irit.fr}
       
\makeatletter
\newenvironment{breakablealgorithm}
  {
   \begin{center}
   \small 
     \refstepcounter{algorithm}
     \hrule height.8pt depth0pt \kern2pt
     \renewcommand{\caption}[2][\relax]{
       {\raggedright\textbf{\ALG@name~\thealgorithm} ##2\par}%
       \kern2pt\hrule\kern2pt
     }
  }{
     \kern2pt\hrule\relax
   \end{center}
  }
  
\makeatother


\begin{abstract}

Using data warehouses to analyse multidimensional data is a significant task in company decision-making. The need for analyzing data stored in different data warehouses generates the requirement of merging them into one integrated data warehouse. The data warehouse merging process is composed of two steps: matching multidimensional components and then merging them. Current approaches do not take all the particularities of multidimensional data warehouses into account, e.g., only merging schemata, but not instances; or not exploiting hierarchies nor fact tables. Thus, in this paper, we propose an automatic merging approach for star schema-modeled data warehouses that works at both the schema and instance levels. We also provide algorithms for merging hierarchies, dimensions and facts. Eventually, we implement our merging algorithms and validate them with the use of both synthetic and benchmark datasets.
\end{abstract}

\maketitle

\section{Introduction}

Data warehouses (DWs) are widely used in companies and organizations as an important Business Intelligence (BI) tool to help build decision support systems \cite{Salvatore07}. Data in DWs are usually modeled in a multidimensional way, which allows users to consult and analyze the aggregated data through multiple analysis axes with On-Line Analysis Processing (OLAP)~\cite{Romero09}. In a company, various independent DWs containing some common elements and data may be built for different geographical regions or functional departments. There may also exist common elements and data between the DWs of different companies. The ability to accurately merge diverse DWs into one integrated DW is therefore considered as a major issue~\cite{Kwakye13}.  DW merging constitutes a promising solution to provide more opportunities of analysing the consistent data coming from different sources.

A DW organizes data according to analysis subjects (facts) associated with analysis axes (dimensions). Each fact is composed of indicators (measures). Finally, each dimension may contain one or several analysis viewpoints (hierarchies). Hierarchies allow users to aggregate the attributes of a dimension at different levels to facilitate analysis. Hierarchies are identified by attributes called parameters.

Merging two DWs is a complex task that implies solving several problems. The first issue is identifying the common basic components (attributes, measures) and defining semantic relationships between these components. The second issue is merging schemata that bear common components. Merging two multidimensional DWs is difficult because two dimensions can (1) be completely identical in terms of schema, but not necessarily in terms of instances; (2) have common hierarchies or have sub-parts of hierarchies in common without necessarily sharing common instances. Likewise, two schemata can deal with the same fact or different facts, and even if they deal with the same facts, they may or may not have measures in common, without necessarily sharing common data.

Moreover, a merged DW should respect the constraints of the input multidimensional elements, especially the hierarchical relationships between attributes. When we merge two dimensions having matched attributes of two DWs, the final DW should preserve all the partial orders of the input hierarchies (i.e., the binary aggregation relationships between parameters) of the two dimensions. It is also necessary to integrate all the instances of the input DWs, which may cause the generation of empty values in the merged DW. Thus, the merging process should also include a proper analysis of empty values.

In sum, the DW merging process concerns matching and merging tasks. The matching task consists in generating correspondences between similar schema elements (dimension attributes and fact measures)  \cite{Philip11} to link two DWs. 
The merging task is more complex and must be carried out at two levels: the schema level and the instance level. Schema merging is the process of integrating several schemata into a common, unified schema \cite{Quix07}. Thus, DW schema merging aims at generating a merged unified multidimensional schema. The instance level merging deals with the integration and management of the instances. In the remainder of this paper, the term ``matching'' designates schema matching without considering instances, while the term ``merging''  refers to the complete merging of schemata and corresponding instances.

To address these issues, we define an automatic approach to merge two DWs modeled as star schemata (i.e., schemata containing only one fact table), which (1) generates an integrated DW conforming to the multidimensional structures of the input DWs, (2) integrates the input DW instances into the integrated DW and copes with empty values generated during the merging process.

The remainder of this paper is organized as follows. In Section 2, we review the related work about matching and merging DWs. In Section 3, we specify an automatic approach to merge different DWs and provide DW merging algorithms at the schema and instance levels. In Section 4, we experimentally validate our approach. Finally, in Section 5, we conclude this paper and discuss  future research.

\section{Related work} 
DW merging actually concerns the matching and the merging of multidimensional elements. 
We classify the existing approaches into four levels: matching   multidimensional components, matching multidimensional schemata, merging  multidimensional schemata and merging  DWs.

A multidimensional component matching approach for matching aggregation levels is 
based on the fact that the cardinality ratio of two aggregation levels from the same hierarchy is nearly always the same, no matter the dimension they belong to \cite{Bergamaschi11}. Thus, by creating and manipulating the cardinality matrix for different dimensions, it is possible to discover the matched attributes.

The matching of multidimensional schemata directs at discovering the matching of every multidimensional components between two multidimensional schemata. A process to automatically match 
two multidimensional schemata is achieved by evaluating the semantic similarity of  multidimensional component names \cite{Banek07}. Attribute and measure data types are also compared in this way. The selection metric of bipartite graph helps determine the mapping choice and define rules aiming at preserving the partial orders of the hierarchies at mapping time. Another approach matches a set of star schemata generated from both business requirements and data sources \cite{Elamin18}. Semantic similarity helps find the matched facts and dimension names. Yet, the DW designer must intervene to manually identify some elements.

A two-phase approach for automatic multidimensional schema merging is achieved by transforming 
the multidimensional schema into a UML class diagram \cite{Jamel05}. Then, class names are compared and the number of common attributes relative to the minimal number of attributes of the two classes is computed to decide whether two classes can be merged.

DW merging must operate at both schema and instance levels. Two DW merging approaches are the intersection and union of the matched dimensions. 
Instance merging is realized by a d-chase procedure \cite{Torlone08}. 
The second merging strategy exploits similar dimensions based on the equivalent levels in schema merging \cite{Olaru12}. It also uses the d-chase algorithm for instance merging. However, the two approaches above 
do not consider the fact table. Another DW merging approach   is based on the lexical similarity of schema string names and instances, and by considering schema data types and constraints \cite{Kwakye13}. Having the mapping correspondences, the merging algorithm takes the preservation requirements of the multidimensional elements into account, and is formulated to build the final consolidated DW. However, merging details are not precise enough and  hierarchies are not considered.

To summarize, none of the existing merging methods can satisfy our DW merging requirements. Some multidimensional components are ignored in these approaches, and the merging details of each specific multidimensional components is not explicit enough, which motivates us to propose a complete DW merging approach.

\section{Preliminaries}
 We introduce in this section the basic concepts of  multidimensional DW design \cite{Ravat08}. The multidimensional DW can be modelled by a star or a constellation schema. In the star schema, there is a single fact connected with different dimensions, while the constellation schema consists of more than one fact which share one or several common dimensions. 

\begin{definition}
A constellation denoted $C$ is defined as ($N^{C}, F^{C},$ $D^{C}, Star^{C}$) where $N^{C}$ is a constellation name, $F^{C} = \{F_1^C,...,F_m^C\}$ is a set of facts, $D^{C} = \{D_1^C,...,D_n^C\}$ is a set of dimensions, $Star^{C}:F^{C} \rightarrow 2^{D^{C}}$ associates each fact to its linked dimensions. A star is a constellation where $F^{C}$ contains a single fact; i.e. $m = 1$.
\end{definition}

A dimension models an analysis axis and is composed of attributes (dimension properties). 
\begin{definition}
A dimension, denoted $D \in D^{C}$ is defined as ($N^D, A^D, H^D, I^D$) where $N^D$ is a dimension name, $A^D = \{ a^D_1,...,a^D_u\}$ $ \cup \{id^D\}$ is a set of attributes, where $id^D$ represents the dimension identifier, which is also the parameter of the lowest level and called the root parameter. $H^D = \{H^D_1,...,H^D_v\}$ is a set of hierarchies, $I^D = \{i^D_1,...,i^D_p\}$ is a set of dimension instances. The value of the instance $i^D_p$ for an attribute $a^D_u$ is annotated as $i^D_p.a^D_u$.
\end{definition}
Dimension attributes (also called parameters) are organised according to one or more
hierarchies. Hierarchies represent a particular vision (perspective) and each parameter represents one data granularity according to which measures could be analysed.

\begin{definition}
A hierarchy of a dimension $D$, denoted $H \in H^D$ is defined as $(N^{H}, Param^{H})$ where $N^{H}$ is a hierarchy name, $Param^{H} = <id^D, p^H_2, ..., p^H_v>$ is an ordered set of dimension attributes, called parameters, which represent useful graduations along the dimensions, $\forall k \in [1...v], p^H_k \in A^D$. The roll up relationship between two parameters can be denoted by $p^H_1 \preceq_H p^H_2$ for the case where $p^H_1$ roll up to $p^H_2$ in $H$. For $Param^{H}$, we have $id^D \preceq_H p^H_1, p^H_1 \preceq_H p^H_2, ... ,p^H_{v-1} \preceq_H p^H_v$. The matching of multidimensional schemata is based on the matching of parameters, the matching relationship between two parameters of two hierarchies $p^{H_1}_i$ and $p^{H_2}_j$ is denoted as $p^{H_1}_i \simeq p^{H_2}_j$.
\end{definition}

A sub-hierarchy is a continuous sub-part of a hierarchy which we call the parent hierarchy of the sub-hierarchy. This concept will be used in our algorithms, but it is not really meaningful. So a sub-hierarchy has the same elements than a hierarchy, but its lowest level is not considered as "$id$". All parameters of a sub-hierarchy are contained in its parent hierarchy and have the same partial orders than those in the parent hierarchy. "Continuous" means that in the parameter set of the parent hierarchy of a sub-hierarchy, between the lowest and highest level parameters of the sub-hierarchy, there is no parameter which is in the parent hierarchy but not in the sub-hierarchy.

\begin{definition}
A sub-hierarchy $SH$ of $H \in H^D$ is defined as $(N^{SH}, Param^{SH})$ where $N^{SH}$ is a sub-hierarchy name, $Param^{SH} =$ $<p^{SH}_1, ..., p^{SH}_v>$ is an ordered set of parameters, called parameters, $\forall k \in [1...v], p^H_k \in Param^{H}$. According to the relationship between a sub-hierarchy and its parent hierarchy, we have: (1) $\forall p^{SH}_1, p^{SH}_2 \in Param^{SH},$ $p^{SH}_1 \preceq_{SH} p^{SH}_2$ $\Rightarrow p^{SH}_1, p^{SH}_2 \in Param^H \land p^{SH}_1 \preceq_H p^{SH}_2$, (2) $\forall p^H_1, p^H_2, p^H_3 \in Param^H,$ $p^H_1 \preceq_H p^H_2 \land p^H_2 \preceq_H p^H_3 \land p^H_1, p^H_3 \in Param^{SH} \Rightarrow p^H_2 \in Param^{SH}$.
    
\end{definition}

A fact reflects information that has to be analysed according to dimensions and is modelled through one or several indicators called measures.

\begin{definition}
A fact, noted $F \in F^{C}$ is defined as $(N^F, M^F, I^F,$ $ IStar^F)$ where $N^F$ is a fact name, $M^F = \{ m^F_1,..., m^F_w \}$  is a set of measures. $I^F = \{i^F_1,..., i^F_q\}$  is a set of fact instances. The value of a measure $m^F_w$ of the instance $i^F_q$ is denoted as $i^F_q.m^F_w$. $IStar^F : I^F \to \mathcal{D}^F$ is a function where $\mathcal{D}^F$ is the cartesian product over sets of dimension instances, which is defined as $\mathcal{D}^F$ = $\prod_{D_k \in Star^{C}(F)} I^{D_k}$. $IStar^F$ associates fact instances to their linked dimension instances. 
\end{definition}
We complete these definitions by a function 
$extend(H_1, H_2)$ allowing to extend the parameters of the first (sub)hierarchy $H_1$ by the other one ($H_2$). 

\section{An automatic approach for DW merging}
\begin{figure}[h] 
\centering
\setlength{\belowcaptionskip}{-6pt}
 \includegraphics[width = 0.8\linewidth]{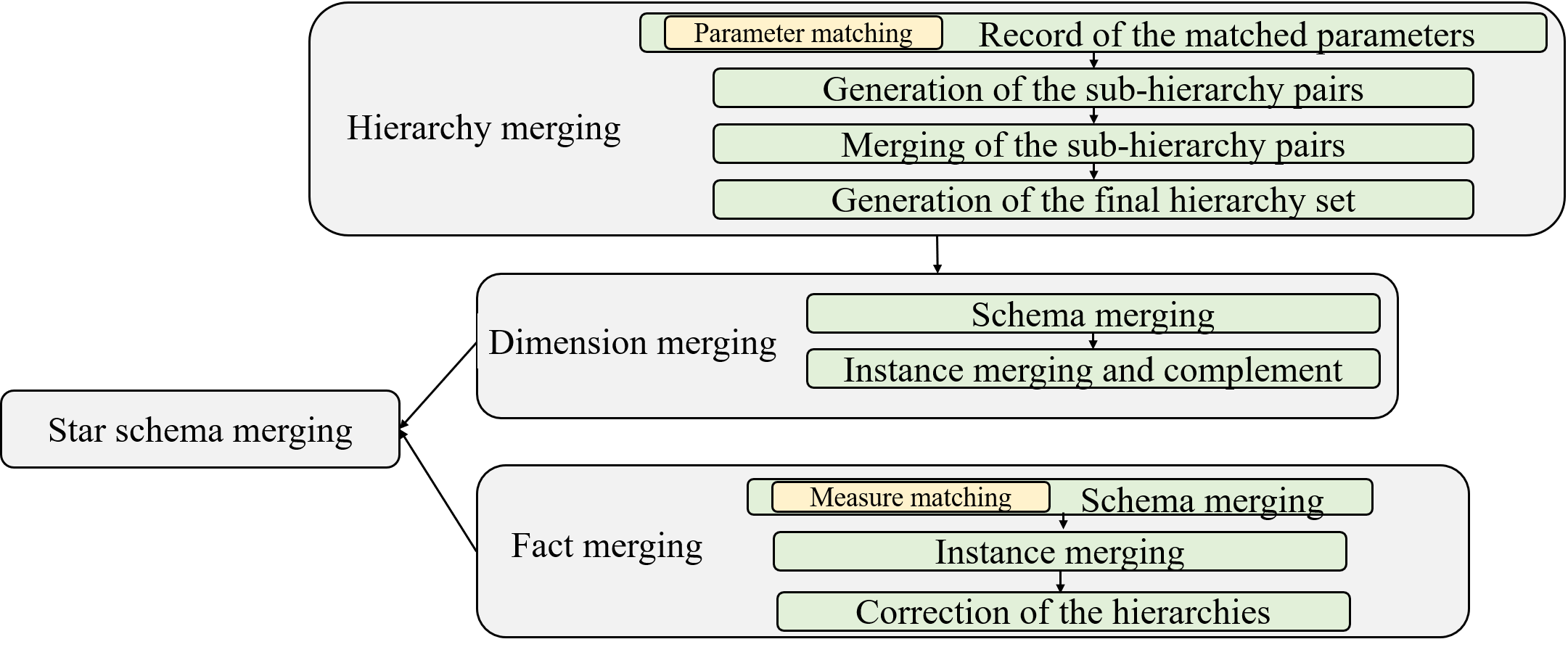}
 \caption{Overview of the merging process}  
 \label{Overview}
\end{figure}

Like illustrated in Figure \ref{Overview}, merging two DWs implies matching steps and steps dedicated to the merging of dimensions and facts. The matching of parameters and measures are based on syntactic and semantic similarities \cite{Meng13}\cite{Elavarasi14} for the attribute or measure names. Since the matching is intensively studied in the literature, we focus in this paper only on the merging steps of our process (green rectangle in Figure \ref{Overview}). In regard to the merging, we firstly define an algorithm for the merging of hierarchies by decomposing two hierarchies into sub-hierarchy pairs and merging them to get the final hierarchy set. Then, we define an algorithm of dimension merging concerning both instance and schema levels and which completes some empty values. Finally, we define an algorithm of the star merging based on the dimension merging algorithm which merges the dimensions and the facts at the schema and instance levels and corrects the hierarchies after the merging.

\subsection{Hierarchy merging}

In this section, we define the schema merging process of two hierarchies coming from two different dimensions. The first challenge is that we should preserve the partial orders of the parameters. The second one is how to decide the partial orders of the parameters coming from different original hierarchies. These challenges are solved in the algorithm proposed below which is achieved by 4 steps: record of the matched parameters, generation of the sub-hierarchy pairs, merging of the sub-hierarchy pairs and generation of the final hierarchy set. 

\begin{breakablealgorithm}
\caption{$MergeHierarchies(H_1, H_2)$}
\hspace{-0.03in} \raggedright {\bf Output:}
A set of merged hierarchies $H'$ or two sets of merged hierarchies $H^{1'}$ and $H^{2'}$ \\
\begin{algorithmic}[1]
\STATE $M, SH', H' \gets \emptyset$; //$M$ is an ordered set of the couples of matched  parameters with possibly the couple of the last parameters, for the $n$th parameter couple $M[n-1]$, $M[n-1][0]$ represents the parameter of $H_1$ in $M[n-1]$, while $M[n-1][1]$ represents the one of $H_2$.
\STATE $Param^{SH_1}, Param^{SH_{1'}}, Param^{SH_2},$ $Param^{SH_{2'}} \gets \emptyset$;
\FOR{\textbf{each} $p^{H_1}_i \in Param^{H_1}$}
    \FOR{\textbf{each} $p^{H_2}_j \in Param^{H_1}$}
        \IF{$p^{H_1}_i \simeq p^{H_2}_j$} 
            \STATE $M \gets M + <p^{H_1}_i,p^{H_2}_j> $;
        \ENDIF
    \ENDFOR
\ENDFOR
\IF{$M = \emptyset$}
    \STATE $H^{1'} \gets \{H_1\}$; $H^{2'} \gets \{H_2\}$;
    \RETURN $H^{1'},H^{2'}$
\ELSE
    \STATE $m_l \gets <Param^{H_1}[|Param^{H_1}|-1], Param^{H_2}[|Param^{H_2}|-1]>$; //pair of the last parameters
    \IF{$m_l \not\in M$}
        \STATE $M \gets M + m_l$;
    \ENDIF
    \FOR{$i = 0$ to $|M|-2$}
        \STATE $p^{SH_1}_1 \gets M[i][0]$; //first parameter of $SH_1$
        \STATE $p^{SH_1}_{v_1} \gets M[i + 1][0]$; //last parameter of $SH_1$
        \STATE $p^{SH_2}_1 \gets M[i][1]$; //first parameter of $SH_2$
        \STATE $p^{SH_2}_{v_2} \gets M[i + 1][1]$; //last parameter of $SH_2$
        \IF{$Param^{SH_1} \subseteq Param^{SH_2}$}
            \STATE $SH' \gets \{SH_2\}$;
        \ELSIF{$Param^{SH_2} \subseteq Param^{SH_1}$}
            \STATE $SH' \gets \{SH_1\}$;
        \ELSIF{$FD_{SH_1\_SH_2} \neq \emptyset$} 
            \FOR{\textbf{each} $Param' \in MergeParameters(FD_{SH_1\_SH_2})$}
                \STATE $Param^{SH_a} \gets Param'$; $SH' \gets SH' + SH_a$;
            \ENDFOR
        \ELSE
            \STATE $SH' \gets \{SH_1, SH_2\}$;
        \ENDIF
        \STATE $H' \gets \{H'_a.extend(SH'_b)|(H'_a \in H') \land  (SH'_b \in SH')\}$;
    \ENDFOR
    \ENDIF
    \IF {$id^{D_1} \simeq id^{D_2}$}
        \STATE $H' \gets H' \cup \{H_1, H_2\}$;
        \RETURN $H'$
    \ELSE
        \STATE $p^{SH_{1'}}_1 \gets p^{H_1}_1$; //first parameter of $SH_{1'}$
        \STATE $p^{SH_{1'}}_{v'_1} \gets M[0][0]$; //last parameter of $SH_{1'}$
        \STATE $p^{SH_{2'}}_1 \gets p^{H_2}_1$; //first parameter of $SH_{2'}$
        \STATE $p^{SH_{2'}}_{v'_2} \gets M[0][1]$; //last parameter of $SH_{2'}$
        \FOR{\textbf{each} $H'_c \in H'$}
            \STATE $H^{1'} \gets SH_{1'}.extend(H'_c )$; $H^{2'} \gets SH_{2'}.extend(H'_c )$;
        \ENDFOR
            \STATE $H^{1'} \gets H^{1'} \cup \{H_1\}$; $H^{1'} \gets H^{2'} \cup \{H_2\}$;
        \RETURN $H^{1'}, H^{2'}$
    \ENDIF
\end{algorithmic}
\end{breakablealgorithm}

\subsubsection{Record of the matched parameters}

The first step of the algorithm consists in matching the parameters of the two hierarchies and record the matched parameter pairs($L_1$-$L_9$). If there is no matched parameter between the two hierarchies, the merging process stops ($L_ {11}$-$L_{12}$).

\subsubsection{Generation of the sub-hierarchy pairs}

Then the algorithm generates pairs containing 2 sub-hierarchies ($SH_1$ and $SH_2$) of the original hierarchies whose lowest and highest level parameters are adjacent in the list of matched parameter pairs that we created in the previous step ($L_{18}$-$L_{22}$). To make sure that the last parameters of the two hierarchies are included in the sub-hierarchies, we also add the pair of the last parameters into the matched parameter pair ($L_{14}$-$L_{17}$).

\begin{figure}[h]
\centering

\setlength{\belowcaptionskip}{-6pt}
 \includegraphics[width = 0.8\linewidth]{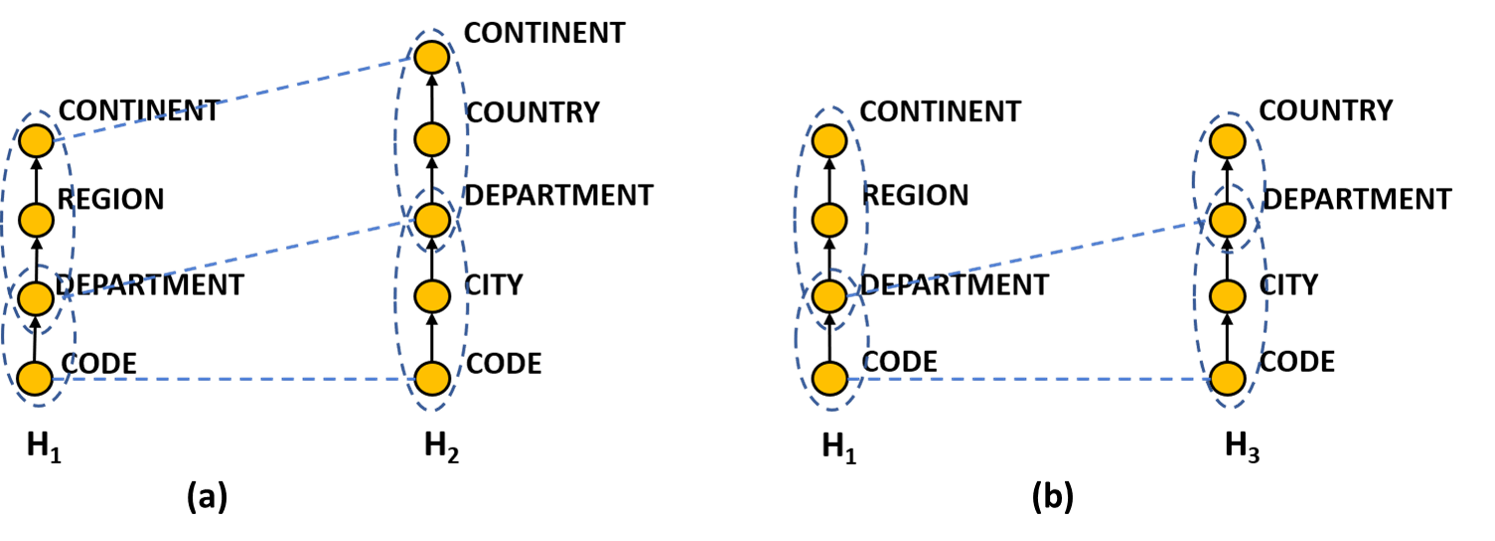}
 \caption{Example of generation of the sub-hierarchy pairs}  
 \label{genesub}
\end{figure}

\begin{example} In Figure \ref{genesub}, for (a), we have $H1.Code \simeq H2.Code$, $H1.Department \simeq H2.Department$, $H1.Continent \simeq H2.Conti$- $nent$. So for the first sub-hierarchy pair, the first parameter of $SH_1$ and $SH_2$ is $Code$ and their last parameter is $Department$, so we have: $Param^{SH_1} = < Code,$ $Department>$, $Param^{SH_2} = <Code, City,$ $Department>$. In the second sub-hierarchy pair, we get the sub-hierarchy of $H_1$ from $Department$ to $Continent$ : $Param^{SH_1}$ $= <Department, Region,$ $Continent>$, and the sub-hierarchy of $H_2$ from $Department$ to $Continent$ : $Param^{SH_2} = <Department, Coun$- $try, Continent>$. If the last parameters of the two original hierarchies do not match, like $Continent$ of $H_1$ and $Country$ of $H_3$ in (b), $<Continent, Country>$ is added into the matched parameter pair $M$ of the algorithm so that the last sub-hierarchies of $H_1$ and $H_3$ are $Param^{SH_1} = <Department, Region,$ $Continent>$ and $Param^{SH_3} = <Department, Country>$.
\end{example}

\subsubsection{Merging of the sub-hierarchies}
We then merge each sub-hierarchy pair to get a set of merged sub-hierarchies ($SH'$) and combine each of these sub-hierarchy sets to get a set of merged hierarchies ($H'$) ($L_{23}$-$L_{35}$).

The matched parameters will be merged into one parameter, so it's the unmatched parameters that we should deal with. We have 2 cases in terms of the unmatched parameters.

If one of the sub-hierarchies has no unmatched parameter, we obtain a sub-hierarchy set containing one sub-hierarchy whose parameter set is the same as the other sub-hierarchy ($L_{23}$-$L_{26}$).

\begin{example} For the first parameter pair $SH_1 = <Code, Depart$- $ment>$ and $SH_2=<Code, City, Department>$ of $H_1$ and $H_2$ in Figure \ref{hiemer1}. We see that $SH_1$ does not have any unmatched parameter, so the obtained sub-hierarchy set contains one sub-hierarchy whose parameter set is the same as $SH2$ which is $Param^{SH'}=<<Code, City, Department>>$.
\end{example}

The second case is that both two sub-hierarchies have unmatched parameters ($L_{27}$-$L_{30}$). We then see if these unmatched parameters can be merged into one or several hierarchies and discover their partial orders. Our solution is based on the functional dependencies (FDs) of these parameters. To be able to detect the FDs of the parameters of the two sub-hierarchies, we should make sure that there are intersections between the instances of these two sub-hierarchies which means that they should have same values on the root parameter of the sub-hierarchies. We keep only the FDs which have a single parameter in both hands and which can not be inferred by transitivity. These FDs are represented in the form of ordered set ($FD_{SH_1\_SH_2}$) are then treated by algorithm 2  $MergeParameters$ to get the parameter sets of the merged sub-hierarchies. If it's not possible to discover the FDs, the two sub-hierarchies are impossible to be merged ($L_{31}$-$L_{32}$).

Algorithm 2  $MergeParameters$ constructs recursively the parameter sets from the FDs in the form of ordered sets. In each recursion loop, for each one of these sets, we search for the other ones whose non-last (or non-first) elements have the same values and order as its non-first (or non-last) elements and then merge them ($L_{6}$-$L_{21}$). The recursion is finished until there are no more two sets being able to be merged ($L_{22}$-$L_{31}$).

\begin{figure}[h]
\centering
\setlength{\belowcaptionskip}{-6pt}
 \includegraphics[width = \linewidth]{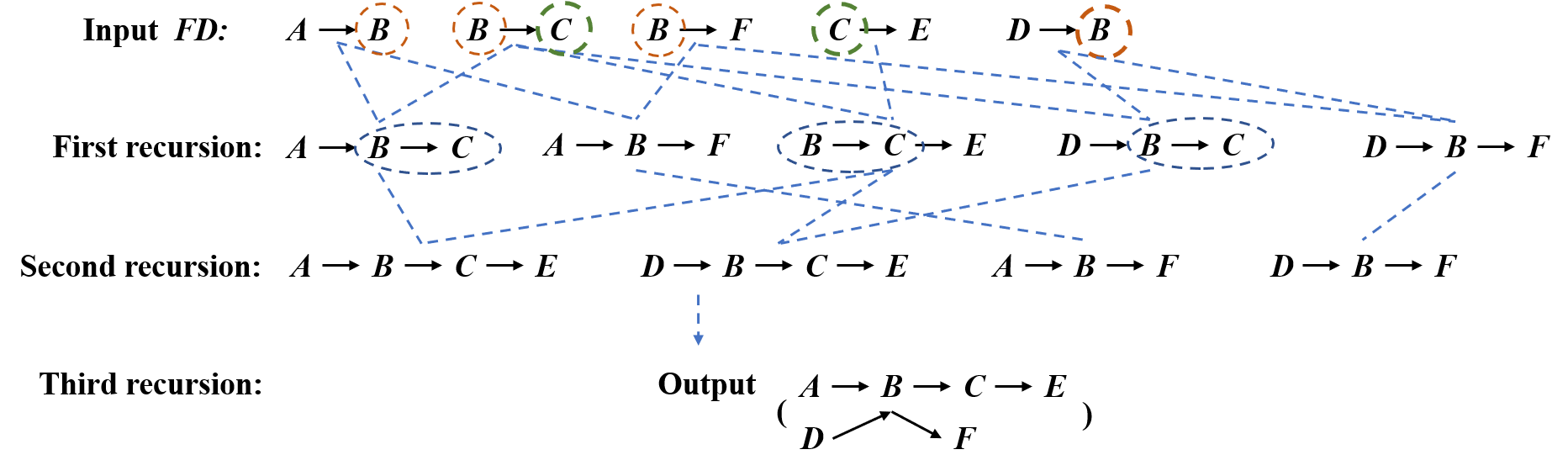}
 \caption{Example of parameter merging based on FDs}  
 \label{parmer}
\end{figure}

\begin{example}If we have $FD = <<A, B>, <B, C>, <B, F>, <C, E>, <D, B>>$. Like illustrated in Figure \ref{parmer}, in the first recursion, by merging the ordered set, we get $Param = <<A, B, C>, <A, B, F>, <B, C, E>, <D, B, C>, <D, B, F>>$, all the ordered sets in $FD$ are merged, so there are only merged ordered set in $Param$. $Param$ is then inputted to the second recursion, we then get the next $Param = <<A, B, C, E>, <D, B, C, E>>$ after the merging of the ordered sets, since $<A, B, F>$ and $<D, B, F>$ are not merged, they are also added into $Param$, and we get $Param = <<A, B, C, E>, <D, B, C, E>, <A, B, F>,  <D, B, F>>$. In the final recursion, it's no more possible to merge any two ordered sets, so the parameter set of the final result of the hierarchy set is $<<A, B, C, E>, <D, B, C, E>, <A, B, F>,  <D, B, F>>$.
\end{example}

\begin{breakablealgorithm}
\caption{$MergeParameters(FD)$}
\hspace{-0.03in} \raggedright {\bf Output:}
A set of parameter sets $Param$\\
\begin{algorithmic}[1]
\STATE $l \gets |FD|$;
\FOR{$n \gets 0$ to $l-1$ }
    \STATE $fdmerged[n] \gets False$; //Boolean indicating whether an element in $FD$ is mergerd
\ENDFOR
\STATE $existmerged \gets False$; //Boolean indicating whether there are elements that are merged in a recursion loop
\FOR{$i \gets 0$ to $l-1$}
    \FOR{$j \gets i+1$ to $l$}
        \IF{$FD[i][1:l-1] = FD[j][0:l-2]$ //$FD[a][b:c]$ represents the ordered set having the values and order from the $b$th element to the $c$th element of $FD[a]$} 
            \STATE $Param^t \gets FD[i][1:l-1]$ ; 
            \STATE $Param^t \gets Param^t + FD[j][l-1]$;
            \STATE $Param \gets Param + Param^t$;
            \STATE $fdmerged[i], fdmerged[j],$ $existmerged \gets True$;
        \ENDIF
        \IF{$FD[i][0:l-2] = FD[j][1:l-1]$}
            \STATE $Param^t \gets FD[j][1:l-1]$ ;
            \STATE $Param^t \gets Param^t + FD[i][l-1]$;
            \STATE $Param \gets Param + Param^t$;
            \STATE $fdmerged[i], fdmerged[j], existmerged \gets True$;
        \ENDIF
    \ENDFOR
\ENDFOR
    \IF{$existmerged = True$}
        \FOR{$m \gets 0$ to $l-1$}
            \IF{$fdmerged[m] = False$}
                \STATE $Param \gets Param + FD[m]$;
            \ENDIF
        \ENDFOR
        \STATE $Param \gets MergeParameters(Param)$;
    \ELSE
        \STATE $Param \gets FD$;
    \ENDIF
\RETURN $Param$
\end{algorithmic}
\end{breakablealgorithm}

After the merging of each sub-hierarchy pair, we extend the final merged hierarchy set by the new merging result ($L_{34}$). 

\subsubsection{Generation of the final hierarchy set}

\begin{figure}[h]
\centering

\setlength{\belowcaptionskip}{-6pt}
 \includegraphics[width = 0.8\linewidth]{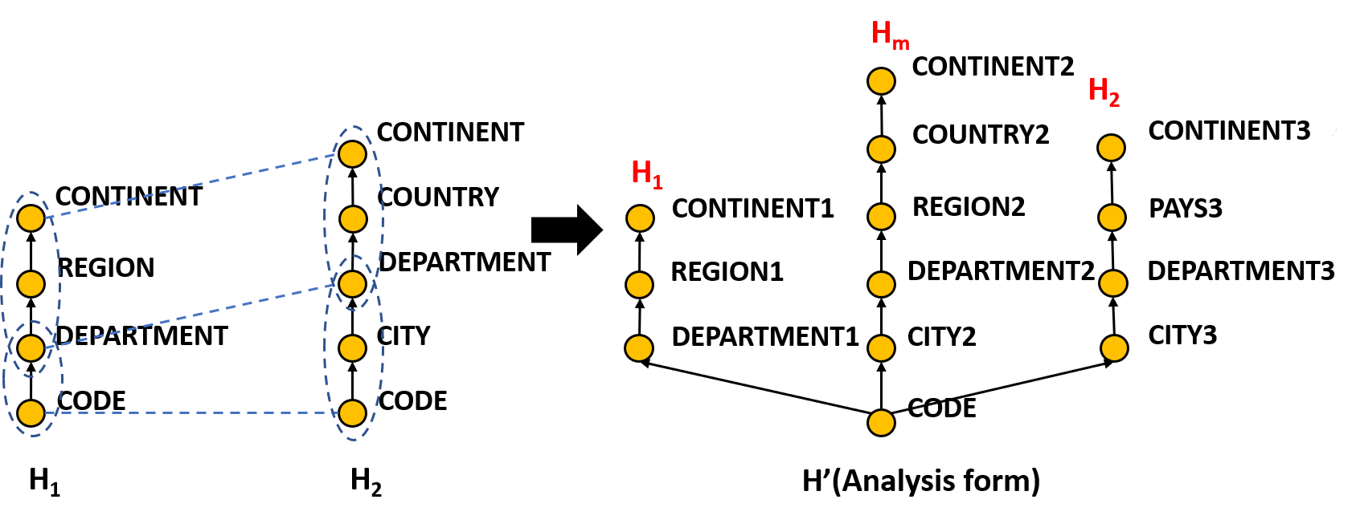}
 \caption{Hierarchy merging example}  
 \label{hiemer1}
\end{figure}

$L_{37}$-$L_{49}$ concerns the generation of the final hierarchy set. The two original hierarchies may have different instances, so there may be empty values in the instances of the merged hierarchies. Some empty values can be completed, which is introduced in the next section of dimension merging. But not all empty values can be completed. The empty values generate the incomplete hierarchies and make the analysis difficult. Inspired by the concept of the structural repair\cite{Sina11}, we also add the two original hierarchies into the final hierarchy set. Then for a parameter which appears in different hierarchies, it can be divided into different parameters in different hierarchies of the hierarchy set so that each hierarchy is complete. Thus, for the multidimensional schema that we get, we provide an analysis form like shown in Figure \ref{hiemer1}. In the analysis form, one parameter can be marked with different numbers if it is in different hierarchies. 

For the generation of the final hierarchy set, we discuss 2 cases where the 2 hierarchies have the matched root parameters which means their dimensions are the same analysis axis and the opposite case which will lead to 2 kinds of output results (one or two sets of merged hierarchies).

If the root parameters of the two original hierarchies match, we simply add the two original hierarchies into the merged hierarchy set obtained in the previous step to get one final merged hierarchy set. ($L_{37}$-$L_{39}$).
\begin{example}
For the hierarchies $H_1$ and $H_2$ in Figure \ref{hiemer1}, we combine the merged hierarchy obtained in $Example$ $4.4$ with the result gained in $Example$ $4.2$ to get the merged hierarchy $H_m$ : $<Code, City, Department, Region, Country, Continent>$. We add $H_m$ into the hierarchy set $H'$ and then also add the original hierarchies $H_1$ and $H_2$. Thus $H'$ is the final merged hierarchy set.
\end{example}
If the root parameters of the two original hierarchies do not match, we will get two merged hierarchy sets instead of one. For each original hierarchy, the final merged hierarchy set will be the extension of the sub-hierarchy containing all the parameters which are not included in any one of the sub-hierarchies created before ($SH_{1'}$ and $SH_{2'}$) with the merged hierarchy set that we get plus this original hierarchy itself ($L_{41}$-$L_{49}$).

\begin{example}
In Figure \ref{dimmer1}, between $H_1$ and $H_3$, we have $H_1.Depart$- $ment \simeq H_3.Department$ and $H_1.Continent \simeq H_3.Continent$. We can then get one sub-hierarchy pair in which there are 2 sub-hierarchies containing parameter sets $<Department, Region, Con$- $tinent>$ and $<Department, Country, Continent>$. By merging the sub-hierarchy pairs, we get the merged hierarchy whose parameter set is $<Department, Region, Country, Continent>$. For $H_1$, the remaining part $<Code>$ is associated to it to get the merged hierarchy $H_{13}^1$. We then get the merged hierarchy set of $H_1$ containing $H_1$ and $H_{13}^1$. We do the same thing for $H_3$ and get the merged hierarchy set containing $H_3$ and $H_{13}^2$. 
\end{example}

\subsection{Dimension merging}
This section concerns the merging of two dimensions having matched attributes which is realized by algorithm 3 $MergeDimensions$. We consider both the schema and instance levels for the merging of dimensions. The schema merging is based on the merging of hierarchies. Concerning the instances, we have 2 tasks: merging the instances and completing the empty values. 

\begin{breakablealgorithm}
\caption{$MergeDimensions(D_1, D_2)$}
\hspace{-0.03in} \raggedright {\bf Output:} One merged dimension $D'$ or two merged dimensions $D^{1'}$ and $D^{2'}$\\
\begin{algorithmic}[1]
\IF{$id^{D_1} \simeq id^{D_2}$}
    \STATE $H^{D'} \gets \emptyset$;
    \FOR{\textbf{each} $H^{D_1}_i \in H^{D_1}$}
            \FOR{\textbf{each} $H^{D_2}_j \in H^{D_2}$ }
                \STATE $H^{D'} \gets H^{D'} \cup MergeHierarchies(H^{D_1}_i, H^{D_2}_j)$;
            \ENDFOR
    \ENDFOR
    \STATE $A^{D'} \gets A^{D_1} \cup A^{D_2}$; $H^m \gets H^{D'} \setminus (H^{D_1} \cup H^{D_2})$;
    \STATE $CompleteEmpty(D', D', H^m)$;
    \RETURN $D'$
\ELSE
    \STATE $H^{D^{1'}}, H^{D^{2'}}, A^{D^{1'}}, A^{D^{2'}} \gets \emptyset$;
    \FOR{\textbf{each} $H^{D_1}_i \in H^{D_1}$}
        \FOR{\textbf{each} $H^{D_2}_j \in H^{D_2}$ }
            \STATE $H^{1'}, H^{2'}\gets$  $MergeHierarchies$($H^{D_1}_i, H^{D_2}_j$);
            \STATE $H^{D^{1'}} \gets H^{D^{1'}} \cup H^{1'}$; $H^{D^{2'}} \gets H^{D^{2'}} \cup H^{2'}$;
        \ENDFOR
    \ENDFOR
    \FOR{\textbf{each} $H^{D^{1'}}_u \in H^{D^{1'}}$}
        \STATE $A^{D^{1'}} \gets A^{D^{1'}} \cup Param^{H^{D^{1'}}_u}$;
    \ENDFOR
    \FOR{\textbf{each} $H^{D^{2'}}_v \in H^{D^{2'}}$}
        \STATE $A^{D^{2'}} \gets A^{D^{2'}} \cup Param^{H^{D^{2'}}_v}$ ;
    \ENDFOR
    \STATE $H^{m_1} \gets H^{D'} \setminus H^{D_1}$; $H^{m_2} \gets H^{D'} \setminus H^{D_2}$;
    \STATE  $CompleteEmpty(D^{1'}, D^{2'}, H^{m_1})$;
    \STATE  $CompleteEmpty(D^{2'}, D^{1'}, H^{m_2})$;
    \RETURN $D^{1'}, D^{2'}$
\ENDIF
\end{algorithmic}
\end{breakablealgorithm}

\subsubsection{Schema merging}
If the root parameters of the two dimensions match, the algorithm generates a merged dimension ($L_{1}$-$L_{8}$). The hierarchy set of the merged dimension is the union of the hierarchy sets generated by merging every 2 hierarchies of the original dimensions ($L_{3}$-$L_{7}$). We also get a hierarchy set containing only the merged hierarchies but no original hierarchies ($H^m$) which is to be used for the complement of the empty values ($L_{8}$). The attribute set of the merged dimension is the union of the attribute sets of the original dimensions ($L_{8}$).

\begin{example}
Given 2 original dimensions $D_1$ and $D_2$ in Figure \ref{Starmer1} and their instances in Figure \ref{dimmeri1}, we can get the merged dimension schema $D'$ in Figure \ref{Starmer1}. In $D'$, $H_1$ and $H_2$ are the original hierarchies of $D_1$, $H_3$ and $H_4$ are those of $D_2$, $H_{13}$ is a merged hierarchy of $H_1$ and $H_3$, and $H_{24}$ is  a merged hierarchy of $H_2$ and $H_4$. We can thus get $H^{D'} = \{H_1, H_2, H_3, H_4, H_{13}, H_{24}\}$, $H_m = \{H_{13}, H_{24}\}$, $A^{D'} = \{Code, City, Department, Region, Country, Continent, Profession,\\Subcategory, Category\}$
\end{example}
\label{exdimm}
\begin{figure}[h]
\centering

\setlength{\belowcaptionskip}{-6pt}
 \includegraphics[width = 0.8\linewidth]{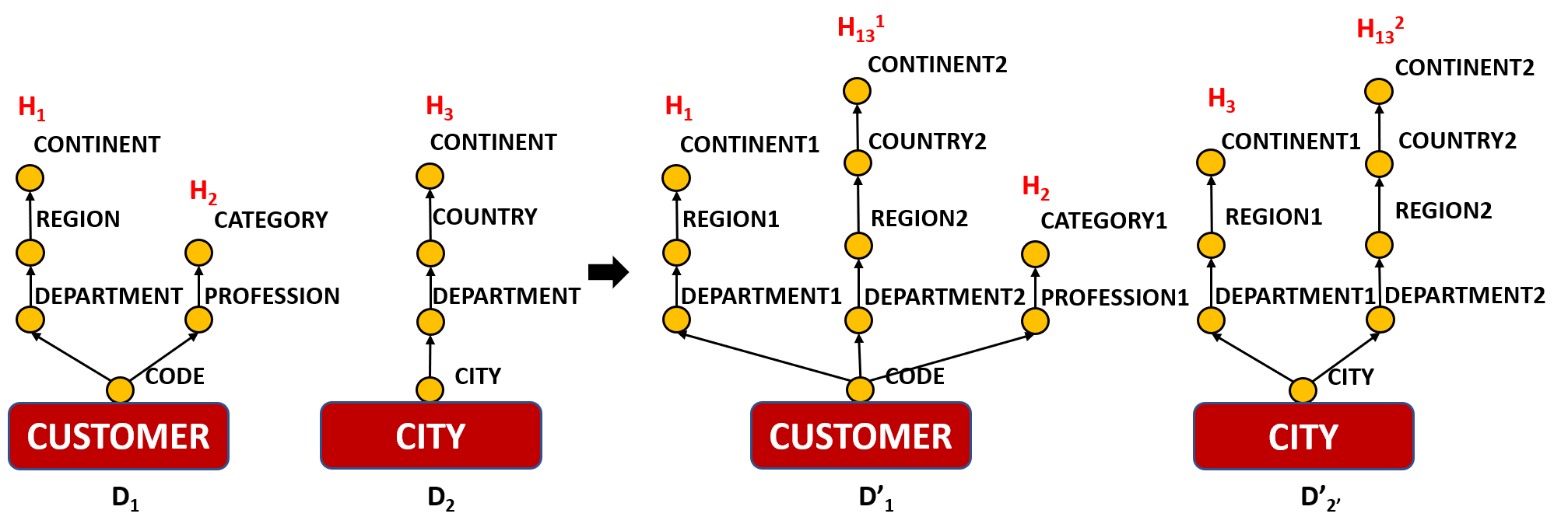}
 \caption{Dimension merging example (schema)}  
 \label{dimmer1}
\end{figure}
When the root parameters of the two dimensions don't match, we will get a merged dimension for each original dimension, which is realized by $L_{13}$-$L_{25}$. For each original dimension, the hierarchy set of its corresponding merged dimension is the union of all hierarchy sets generated by merging every 2 hierarchies of the original dimensions ($L_{13}$-$L_{18}$), the attribute set is the union of the attributes of each hierarchy in the merged dimension ($L_{19}$-$L_{24}$). Similar to the first case, we get a hierarchy set containing only the merged hierarchies for each original dimension ($H^{m_1}$ and $H^{m_2}$) ($L_{26}$-$L_{27}$).

\begin{example}
Given 2 original dimensions $D_1$ and $D_2$ in Figure \ref{dimmer1} and their instances in Figure \ref{dimmeri2}, after the execution of algorithm 3 $MergeDimensions$, we can get the merged dimension schema $D^{1'}$ and $D^{2'}$ in Figure \ref{dimmer1}. In $D^{1'}$, $H_1$ and $H_2$ are the original hierarchies of $D_1$, $H_{13}^1$ is the merged hierarchy of $H_1$ and $H_3$. In $D^{2'}$, $H_3$ is the original hierarchy of $D_2$, $H_{13}^2$ is the merged hierarchy of $H_1$ and $H_3$. So for $D_1$, we have $H^{D^{1'}} = \{H_1, H_2, H_{13}^1\}$, $H_{m1} = \{H_{13}^1\}$, $A^{D^{1'}} = \{Code,  Department, Region, Country, Continent, Profession, Cate$-\\$gory\}$, while for $D_2$, we get $H^{D^{2'}} = \{H_3, H_{13}^2\}$, $H_{m2} = \{H_{13}^2\}$, $A^{D^{2'}} = \{City,  Department, Region, Country, Continent\}$
\end{example}

\subsubsection{Instance merging and complement}
\begin{figure}[h]
\centering

\setlength{\belowcaptionskip}{-6pt}
 \includegraphics[width = \linewidth]{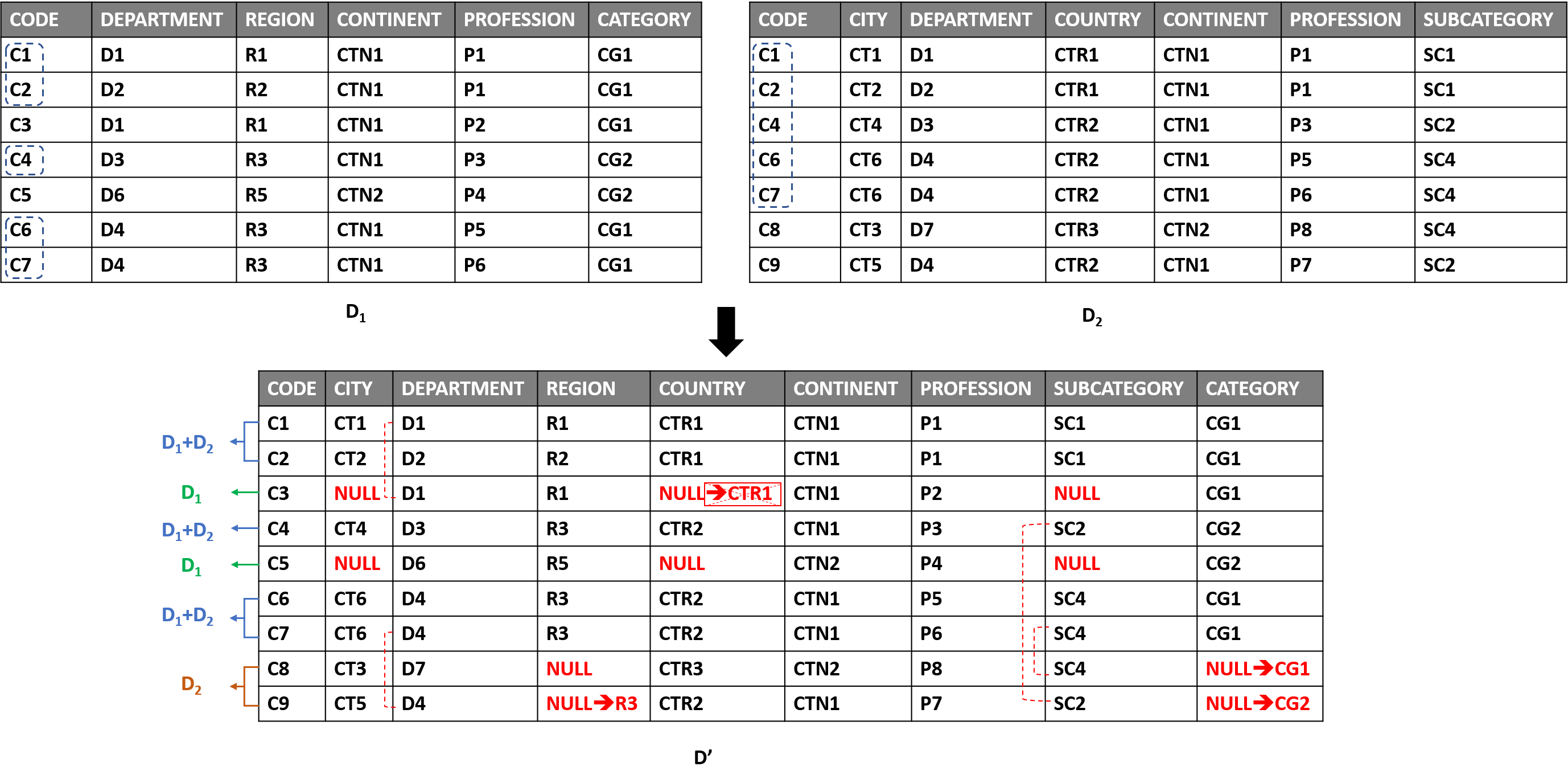}
 \caption{Dimension merging example (instance)}  
 \label{dimmeri1}
\end{figure}

When the root parameters of the two dimensions match, the instance of the merged dimension is obtained by the union of the two original dimension instances which means that we insert the data of the two original dimension tables into the merged dimension table and merge the lines which have the same root parameter instance ($L_{9}$). 
\begin{example}
The instance merging result of Example \ref{exdimm} is presented in Figure \ref{dimmeri1}. All the data in the original dimension tables $D_1$, $D_2$ are integrated into the merged dimension table $D'$. The original tables of the instances are marked on the left of the merged table $D'$ with different colors. There are instances coming from both $D_1$ and $D_2$, which means that they have the same root parameter in $D_1$ and $D_2$, and are therefore merged together.
\end{example}

The attribute set of the merged dimension contains all the attributes of two original dimensions, while the original dimensions may contain their unique attributes. So there may be empty values in the merged dimension table on the instances coming from only one of the original dimension tables and we should complete the empty values on the basis of the existing data ($L_{9}$). 

The complement of the empty values is realized by Algorithm $CompleteEmpty$ where the input $D^{1'}$ is the merged dimension table having empty values to be completed, $D^{2'}$ is the merged dimension table which provides the completed values and $H_m$ is the hierarchy set of $D^{1'}$ containing only merged hierarchies but no original hierarchies. In this discussed case, $D'$ is inputted as both $D^{1'}$ and $D^{2'}$ in $CompleteEmpty$ since we get one merged dimension including all data of two original dimensions ($L_{11}$).

\begin{breakablealgorithm}
\caption{$CompleteEmpty(D^{1'}, D^{2'}, H^m)$}
\begin{algorithmic}[1]
\FOR{\textbf{each} $H^m_a \in H^m$}
    \STATE $I^n \gets \emptyset$;
    \STATE $I^n \gets I^n \cup \{i^{D^{1'}}_k \in I^{D^{1'}}|(i^{D^{1'}}_k.p^{H^m_a}_{1}$ is not null) $\land (\exists p^{H^m_a}_v \in$
    
    $ Param^{H^m_a}, i^{D^{1'}}_k.p^{H^m_a}_v$ is null) $\}$;
    \FOR{\textbf{each} $i^n_b \in I^n$}
        \STATE $P^n \gets \{p^{H^m_a}_v \in Param^{H^m_a}|i^n_b.p^{H^m_a}_v $ is null$\}$;
        \STATE $P^r \gets \{p^{H^m_a}_v \in Param^{H^m_a}|(i^n_b.p^{H^m_a}_v $ is not null) $\land (\forall p^n_s \in P^n, p^{H^m_a}_v \preceq_H p^n_s)\}$;
        \IF{$\exists i^{D^{2'}}_u \in I^{D^{2'}}$ $\exists p^r_w \in P^r, (i^{D^{2'}}_u.p^r_w = i^n_b.p^r_w) \land (\forall p^n_q \in
        P^n, i^{D^{2'}}_u.p^n_q$ is not null)}
            \FOR{\textbf{each} $p^n_c \in P^n$}
                \STATE $i^n_b.p^n_c \gets i^{D^{2'}}_u.p^n_c$;
            \ENDFOR
        \ENDIF
    \ENDFOR
\ENDFOR
\end{algorithmic}
\end{breakablealgorithm}

For an empty value, we search for an instance which has the same value as the instance of this empty value on one of the parameters rolling up to the parameter of the empty value and whose value of the parameter of the empty value is not empty, we can then fill the empty by this non-empty value. The complement of the empty values is also possibly a change of hierarchies. Nevertheless, after completing the empty values of an instance, there may be some completed parameters which are not included in the hierarchies of the instance, so the complement of such values does not make sense in this case. The possible change of the hierarchy is from the hierarchies containing less parameters to those containing more parameters. We know that the merged hierarchies contain more parameters than their corresponding original hierarchies. Hence, before the complement of an instance, we will first look at the merged hierarchies to decide which parameter values can be completed.

In algorithm 4 $CompleteEmpty$ which aims to complete the empty values, for each hierarchy in the merged hierarchy set we see, if \textbf{(a)} there exists instances 
in the merged dimension table which contains empty values on the parameters of this hierarchy ($L_{3}$) and \textbf{(b)} where the value of the second lowest parameter is not empty ($L_{3}$). The condition a is basic because we need empty values to be completed. Since we will complete the empty values by the other lines of the merged dimension table, we can only complete the empty values based on the non-id parameters since the id is unique, so if the second lowest parameter is empty, it can never be completed so that the hierarchy can never be completed. That's why we have the condition b. For each one of the instances satisfying these conditions ($I^n$), we search for the parameters ($P^n$) having empty values ($L_{5}$) and to make sure that each one of them can be completed, we search also for the parameters ($P^r$) which roll up to the lowest of them and to which we refer to complete the empty values ($L_{6}$). We can then complete the empty values like discussed in the previous paragraph ($L_{7}$-$L_{11}$).

\begin{example}
After the merging in $Example$ $4.9$, we get the empty values of $D'$ which are in red in Figure \ref{dimmeri1}. The merged hierarchies are $H_{13}$ and $H_{24}$ as illustrated in Figure \ref{dimmer1}. For $H_{13}$, the instances of code $C3$ and $C5$ have empty values on the second root parameter $City$, which do not satisfy the condition b. As we can see, for the instance of $C3$, although the value of $Country$ can be retrieved through the value of $Department$ which is the same as the instance of $C1$, the value of $City$ can not be completed and thus we should give up this complement. For the instance of $C9$, the value of $Region$ is completed by $C7$ which has the same value of $Department$ and whose value of $Region$ is not empty. When it's the turn of $H_{24}$, values of $Category$ of $C8$ and $C9$ are completed in the same way.
\end{example}

\begin{figure}[h]
\centering

\setlength{\belowcaptionskip}{-6pt}
 \includegraphics[width = \linewidth]{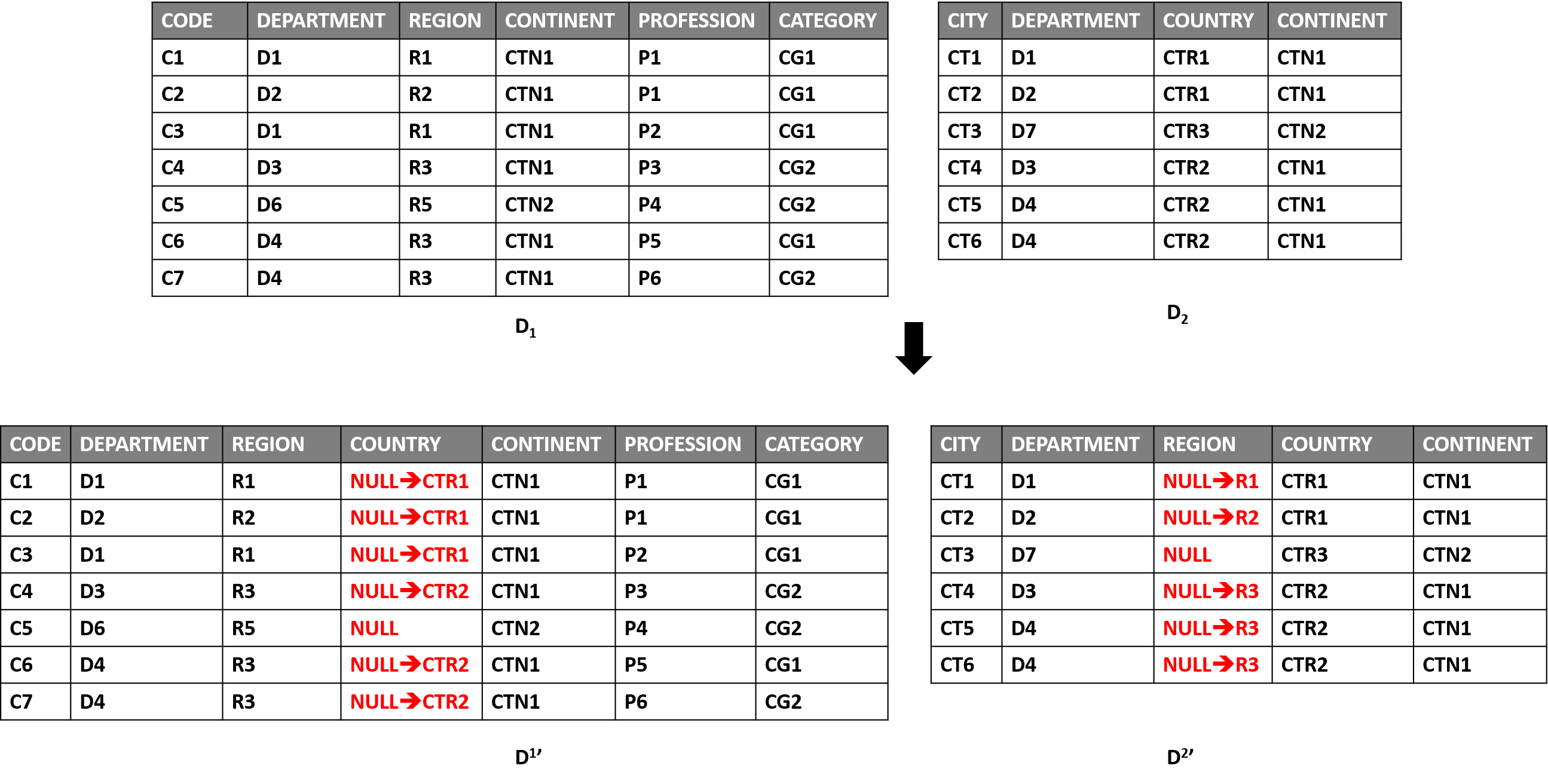}
 \caption{Dimension merging example (instance)}  \label{dimmeri2}
\end{figure}

When the root parameters of the two dimensions don't match, the instance merging and complement are done by $L_{26}$-$L_{27}$. The values of the attributes of one of the dimension tables coming from the other dimension table are empty, so there is only instance complement but no merging. We also call algorithm 4 $CompleteEmpty$ to complete the instances for each one of the merged dimension tables.
\begin{example}
The instance merging and complement of the example for $Example$ $4.8$ is demonstrated in Figure \ref{dimmeri2}. For $D^{1'}$, $Country$ comes from the dimension table $D_2$, so the values of $Country$ are completed by the values in $D^{2'}$. The same operation is also done for $Region$ of $D^{2'}$.
\end{example}

\subsection{Star merging}
In this section, we discuss the merging of two stars. Having two stars, we can get a star schema or a constellation schema because the fact table of each schema may be merged into one schema or not. The star merging is related to the dimension merging and fact merging. Two stars are possible to be merged only if there are dimensions having matched root parameters between them. 

\begin{breakablealgorithm}
\caption{$MergeAllDimensions(S_1, S_2)$}
\hspace{-0.03in} \raggedright {\bf Output:}A set of merged dimensions $D^{S'}$

\begin{algorithmic}[1]
\FOR{\textbf{each} $D^{S_1}_i \in D^{S_1}$}
    \FOR{\textbf{each} $D^{S_2}_j \in D^{S_2}$}
        \IF{${id}^{D^{S_1}_i} \not\simeq {id}^{D^{S_2}_j}$}
            \STATE $D^{S_1}_i, D^{S_2}_j \gets MergeDimensions(D^{S_1}_i, D^{S_2}_j)$;
        \ENDIF
    \ENDFOR
\ENDFOR
\STATE $D^{S'} \gets \emptyset$;
\FOR{\textbf{each} $D^{S_1}_u \in D^{S_1}$}
    \FOR{\textbf{each} $D^{S_2}_v \in D^{S_2}$}
        \IF{${id}^{D^{S_1}_u} \simeq {id}^{D^{S_2}_v}$}
            \STATE $D^{S'} \gets D^{S'} \cup MergeDimensions(D^{S_1}_u, D^{S_2}_v)$;
        \ENDIF
    \ENDFOR
\ENDFOR
\FOR{\textbf{each} $D^{S'}_k \in D^{S'}$}
    \FOR{\textbf{each} $H^{D^{S'}_k}_m \in H^{D^{S'}_k}$}
        \IF{$\not\exists i^{D^{S'}_k}_r \in I^{D^{S'}_k}, (i^{D^{S'}_k}_r$ is on $H^{D^{S'}_k}_m) \lor (i^{D^{S'}_k}_r$ is only on $H^{D^{S'}_k}_m \land (H^{D^{S'}_k}_m \in H^{D^{S_1}} \lor H^{D^{S'}_k}_m \in H^{D^{S_2}}))$}
            \STATE $H^{D^{S'}_k} \gets H^{D^{S'}_k} - H^{D^{S'}_k}_m$;
        \ENDIF
    \ENDFOR
\ENDFOR
\RETURN $D^{S'}$
\end{algorithmic}
\end{breakablealgorithm}

For the dimensions of the two stars, we have two cases: 1. The two stars have the same number of dimensions and for each dimension of one schema, there is a dimension having matched root parameters in the other schema. 2. There exists at least one dimension between the two stars which does not have a dimension having a matched root parameter in the other.

The dimension merging of two stars is common for the two cases which is done by algorithm 5 $MergeAllDimension$. We first merge every two dimensions of the two stars which have unmatched root parameters because the merging of such dimensions is able to complete the original dimensions with complementary attributes ($L_{1}$-$L_{7}$). Then the dimensions having matched root parameters are merged to generate the merged dimensions of the merged multidimensional schema ($L_{8}$-$L_{15}$).
After the merging and complement of the instances of the dimension tables, there may be some merged hierarchies to which none of the instances belong. In this case, if there will be no more update of the data, such hierarchies should be deleted. There may also be original hierarchies in the merged dimensions such that there is no instance which belongs to them but does not belong to any merged hierarchy containing all the parameters of this original hierarchy. The instances belonging to this kind of hierarchies belong also to other hierarchies which contains more parameters,so they become useless and should also be deleted ($L_{18}$-$L_{19}$).

\begin{figure}[h]
\centering

\setlength{\belowcaptionskip}{-6pt}
 \includegraphics[width = 0.8\linewidth]{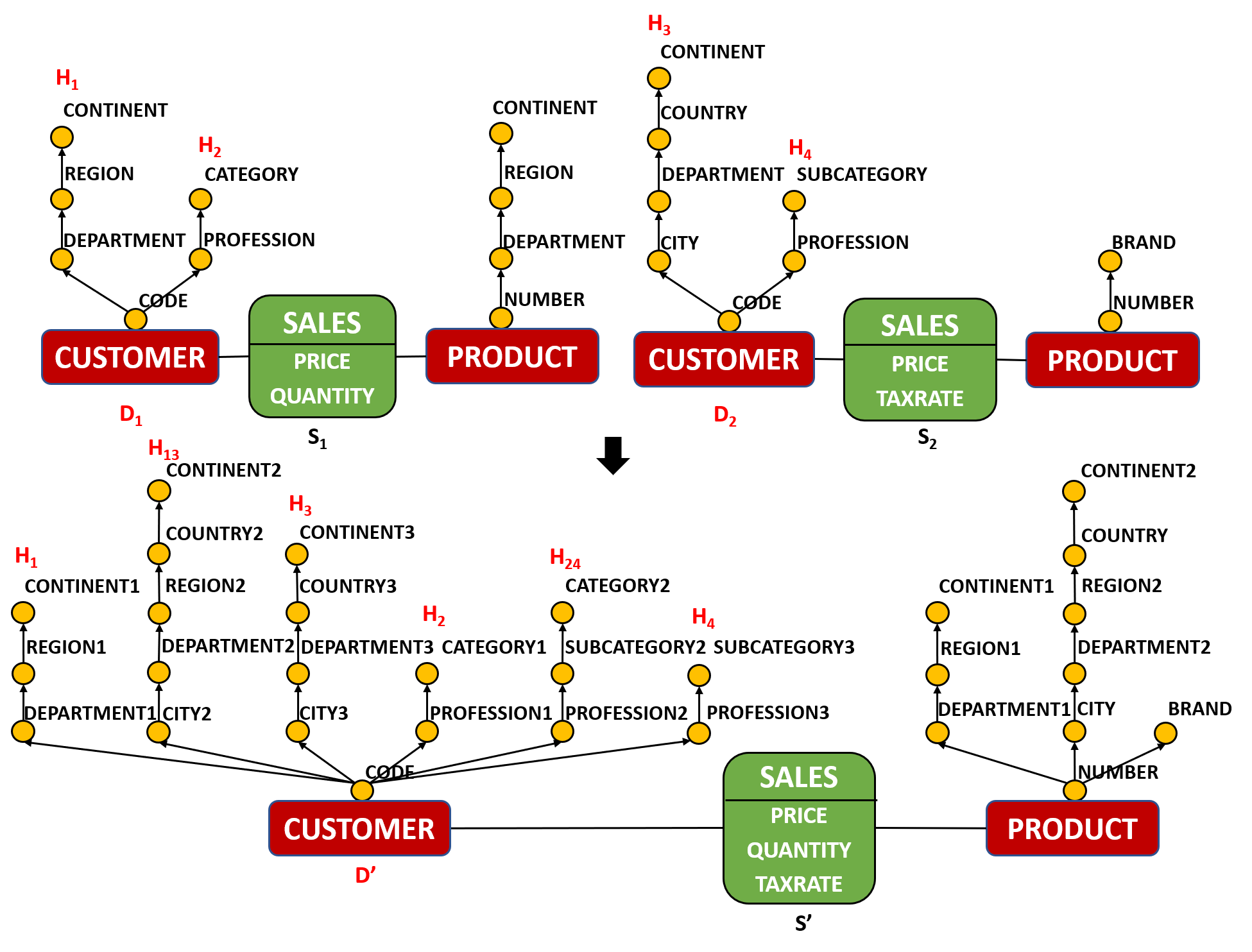}
 \caption{Star merging example (schema)}  
 \label{Starmer1}
\end{figure}

\begin{example}
For the merging of the dimensions of two stars $S_1$ and $S_2$ in Figure \ref{Starmer1}. The dimension $Product$ of $S_1$ and the dimension $Customer$ of $S_2$ are firstly merged since their root parameters don't match but they have other matched parameters. There are then attributes of dimension $Customer$ of $S_2$ added into dimension $Product$ of $S_1$. The two dimensions $Customer$ and the two dimensions $product$ have matched root parameters, so they are merged into the final star schema. After the merging and complement of the instance, we verify each hierarchy in the merged dimension tables. If the merging of $S_1.Customer$ and $S_2.Customer$ is as shown in Figure \ref{dimmer1} at the schema level and in Figure \ref{dimmeri1} at the instance level. In their merged dimension table $D'$. We can find that all the instances belonging to $H_4$ also belong to $H_{24}$ which is a merged hierarchy containing all the parameters of $H_4$, so $H_4$ should be deleted.
\end{example}

We then discuss the merging of the other elements in the two cases which is processed by algorithm 6 $MergeStar$:
  
\begin{breakablealgorithm}
\caption{$MergeStar(S_1, S_2)$}
\hspace{-0.03in} \raggedright {\bf Output:}A merged multidimensional schema which may be a star schema $S'$ or a merged constellation schema $C'$

\begin{algorithmic}[1]
\IF{$(|D^{S_1}| = |D^{S_1}|) \land (\forall D^{S_1}_i \in D^{S_1} \exists D^{S_2}_j \in D^{S_2}, {id}^{D^{S_1}_i} \simeq {id}^{D^{S_2}_j})$}
    \STATE $D^{S'} \gets MergeAllDimensions(S_1, S_2)$;
    \STATE $M^{F^{S'}} \gets M^{F^{S_1}} \cup M^{F^{S_2}}$; $I^{F^{S'}} \gets I^{F^{S_1}} \cup I^{F^{S_2}}$; 
    \STATE $IStar^{F^{S'}} \gets IStar^{F^{S_1}} \cup IStar^{F^{S_2}}$;
    \RETURN $S'$
\ELSE
    \STATE $D^{S'} \gets MergeAllDimensions(S_1, S_2)$; $F^{C'} \gets \{ F^{S'_1}, F^{S'_2}\}$;
    \RETURN $C'$
\ENDIF
\end{algorithmic}
\end{breakablealgorithm}

For the first case, we merge the two fact tables into one fact table and get a star schema. The measure set of the merged star schema is the union of the 2 original measures ($L_{3}$). The fact instances are the union of the measure instances of the two input star schemata ($L_{4}$). The function associating fact instances to their linked dimension instances of the merged schema is also the union of the functions of the original schemata ($L_{4}$).
\begin{figure}[h]
\centering
\setlength{\belowcaptionskip}{-6pt}
 \includegraphics[width = \linewidth]{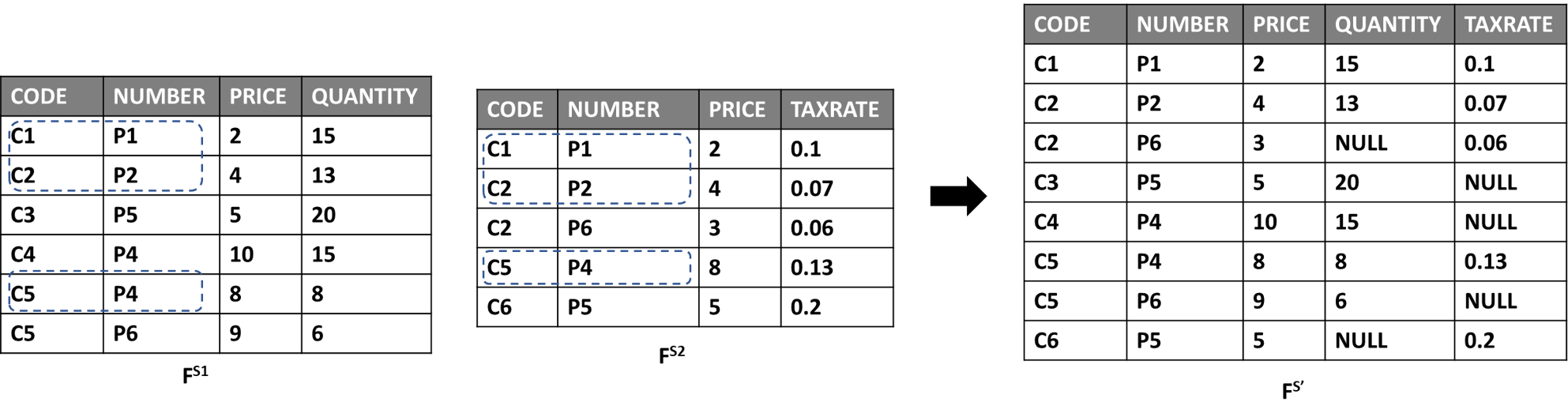}
 \caption{Star merging example (instance)}  
 \label{Starmeri1}
\end{figure}
\begin{example}
For the two original star schemata in Figure \ref{Starmer1}, the dimension merging is discussed above so we mainly focus on the merging of fact table instances here. The dimensions $Customer$, $Product$ of $S_1$ have respectively matched root parameters in the dimensions $Customer$, $Product$ of $S_2$. They also have the same number of dimensions. Therefore we get a merged star schema $S'$, the original fact tables are merged by merging the measures of $S_1$ and $S_2$ to get the fact table of $S'$. At the instance level, in Figure \ref{Starmeri1}, we have the instances of the fact tables, for the instances of $F^{S_1}$ and $F^{S_2}$, the framed parts are the instances having the common linked dimension instances, so they are merged into the merged fact table $F^{S'}$, the other instances are also integrated in $F^{S'}$ but with empty values in the merged instances, but they will not have big impacts on the analysis, so they will not be treated particularly.
\end{example}

For the second case, since there are unmatched dimensions, the merged schema should be a constellation schema. The facts of the original schemata have no change at both the schema and instance levels and compose the final constellation. ($L_{8}$)

\begin{figure}[h]
\centering

\setlength{\belowcaptionskip}{-6pt}
 \includegraphics[width = 0.8\linewidth]{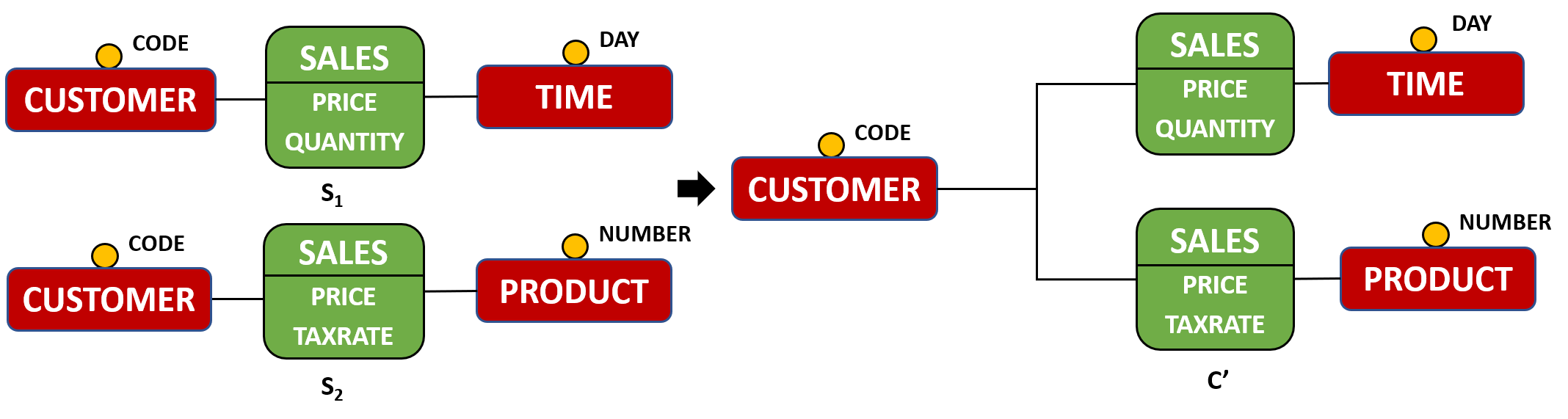}
 \caption{Star merging example (schema)}  
 \label{Starmer2}
\end{figure}

\begin{example}
This example is simplified in Figure \ref{Starmer2} due to the space limit. For the original star schemata $S_1$ and $S_2$, they have dimensions $Customer$ which have the matched root parameters. They also have their unique dimensions: $Time$ of $S_1$ and $Product$ of $S_2$. So the merged schema is a constellation schema generated by merging the dimensions $Customer$ and by keeping the other dimensions and fact tables. At the instance level, we just have a new merged dimension table of $Customer$, the other dimension and fact tables remain unchanged.
\end{example}

\section{Experimental assessments}
To validate the effectiveness of our approach, we applied our algorithms on benchmark data. Unfortunately, we did not find a suitable benchmark for our problem. So, we adapted the datasets of the TPC-H benchmark to generate different DWs. Originally, the TPC-H benchmark serves for benchmarking decision support systems by examining the execution of queries on large volumes of data. Because of space limit, we put the test results in github\footnote{\href{https://github.com/Implementation111/Multidimensional-DW-merging}{https://github.com/Implementation111/Multidimensional-DW-merging}}. 

\subsection{Technical environment and Datasets}
The algorithms were implemented by Python 3.7 and were executed on a processor of Intel(R) Core(TM) i5-8265U CPU@ 1.60GHz with a 16G RAM. The data are implemented in R-OLAP format through the Oracle 11g DBMS. The TPC-H benchmark provides a pre-defined relational schema\footnote{\href{http://tpc.org/tpc\_documents\_current\_versions/pdf/tpc-h\_v2.18.0.png}{http://tpc.org/tpc\_documents\_current\_versions/pdf/tpc-h\_v2.18.0.png}} with 8 tables and a generator of massive data. 

First, we generated 100M of data files, there are respectively 600572, 15000, 25, 150000, 20000, 80000, 5, 1000 tuples in the table of $Lineitem$, $Customer$, $Nation$, $Orders$, $Part$, $Partsupp$, $Region$ and $Supplier$. Second, to have more deeper hierarchies, we included the data of $Nation$ and $Region$ into $Customer$ and $Supplier$, and those of $Partsupp$ into $Part$. Third, we transformed these files to generate two use cases by creating 2 DWs for each case. To make sure that there are both common and different instances in different DWs, for each dimension, instead of selecting all the corresponding data, we selected randomly 3/4 of them. For the fact table, we selected the measures related to these dimension data. Since the methods in the related work do not have exactly the same treated components or objective as the ours, we do not have comparable baseline in our experiments.

\subsection{Star schema generation}
\begin{figure}[h]
\centering

\setlength{\belowcaptionskip}{-6pt}
 \includegraphics[width = 0.8\linewidth]{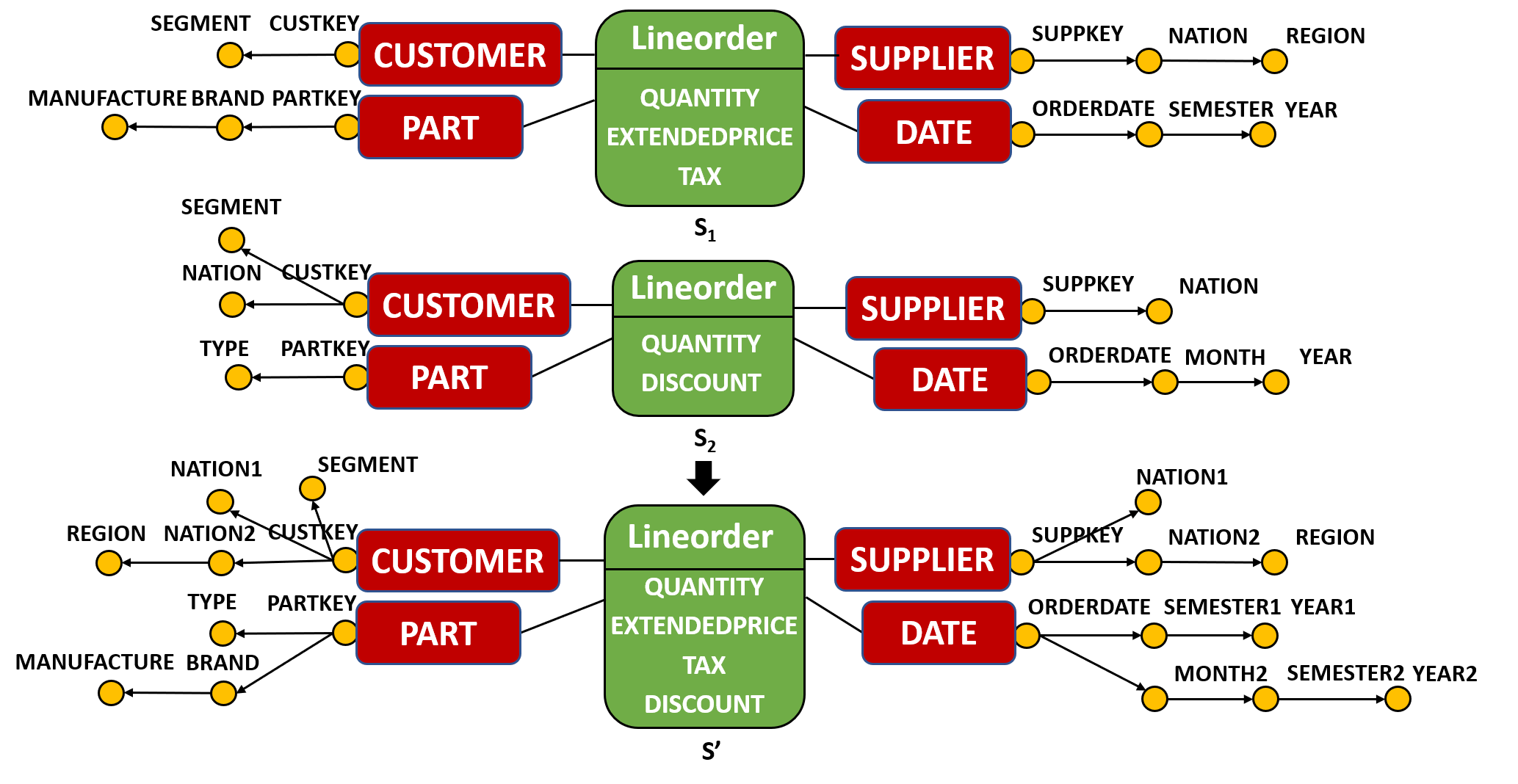}
 \caption{Star schema generation}
  \label{test1}
\end{figure}

The objective of this experiment is to merge two star schemata having the same 4 dimensions 
with the matched lowest level of granularity for each dimension.

After executing our algorithms, we obtain one star schema as shown in Figure \ref{test1} which is consistent with the expectations. The parameters of the hierarchies satisfy the relationships of functional dependency. The run time is 30.70s. The 3 dimensions $Supplier$, $Part$, $Date$ of the original DWs are merged. Between the different dimensions $S_1.Supplier$ and $S_2.Customer$, there is a matched attribute $Nation$, so they are also merged such that $S_1.Supplier$ provides $S_2.Customer$ with the attribute $Region$. Then the $Customer$ in the merged DW also has the attribute $Region$. 
We can also observe that normally, in the merged schema, there should be the original hierarchy  $Orderdate \rightarrow Month \rightarrow Year$ of $S_2.Date$ but which is deleted. By looking up in the table, we find that there is no tuple which belongs to this hierarchy but not to $Orderdate \rightarrow Month \rightarrow Semester \rightarrow year$, that’s why it is removed.

\begin{table}
\small
\centering
\setlength{\tabcolsep}{3.5pt}
\begin{tabular}{ |c|c|c|c|c|c| } 
\hline
 & \textbf{Customer} & \textbf{Supplier} & \textbf{Part} & \textbf{Orderdate} & \textbf{Lineorder}\\
\hline
\textbf{$N_1$} & 11250 & 750 & 15000 & 1804 & 252689\\
\textbf{$N_2$} & 11250 & 750 & 15000 & 1804 & 252821\\
\textbf{$N_{\cap}$} & 8439 & 556 & 11261 & 1349 & 105345\\
\textbf{$N'$} & 14061 & 944 & 18739 & 2259 & 400165\\
\hline
\end{tabular}
    \caption{Number of tuples}
    \label{tab:numtup1}
\end{table}

\begin{table}
\small
\centering
\setlength{\tabcolsep}{0.5pt}
\begin{tabular}{ |c|c|c|c| } 
\hline
 & \textbf{Customer.Region} & \textbf{Supplier.Region} & \textbf{Orderdate.Semester}\\
\hline
\textbf{$N_1$} & X & X & 1804 \\
\textbf{$N_2$} & X & 750 & X \\
\textbf{$N'$} & 9713 & 846 & 2259 \\
\textbf{$N_+$} & 9713 & 96 & 455 \\
\hline
\end{tabular}
    \caption{Number of attributes}
    \label{tab:numattr1}
\end{table}

At the instance level, the result is shown in github. Table \ref{tab:numtup1} shows the number of tuples of the original DWs ($N_1$, $N_2$), of the merged DW ($N'$) and the number of the common tuples ($N_{\cap}$) (tuples having the same dimension key in the original DWs). For each dimension or fact table, $N' = N_1 + N_2 - N_{\cap}$, we can thus confirm that there is no addition or loss of data. For each tuple in the original tables, we verify that the all the values are the same with the values in the merged table. We also find that there are some empty values of the attribute $Region$ in the dimension $Customer$ and $Supplier$ and the attribute $Semester$ of the dimension $Orderdate$ which are completed. Table \ref{tab:numattr1} shows the number of these attributes in the original DWs ($N_1$, $N_2$) and in the merged DW ($N'$), we can then get the number of the completed values $N_+$ for these attributes. They meet the relationship $N’ = N_1 + N_2 + N_+$.

\subsection{Constellation schema generation}

The objective of this experiment is to merge two star schemata having the same 2 dimensions ($Customer$, $Supplier$) with the same lowest level of granularity for each dimension, as well as 2 different dimensions ($S_1.Part$ and $S2.Date$).

\begin{figure}[h]
\centering
\setlength{\belowcaptionskip}{-6pt}
 \includegraphics[width = 0.8\linewidth]{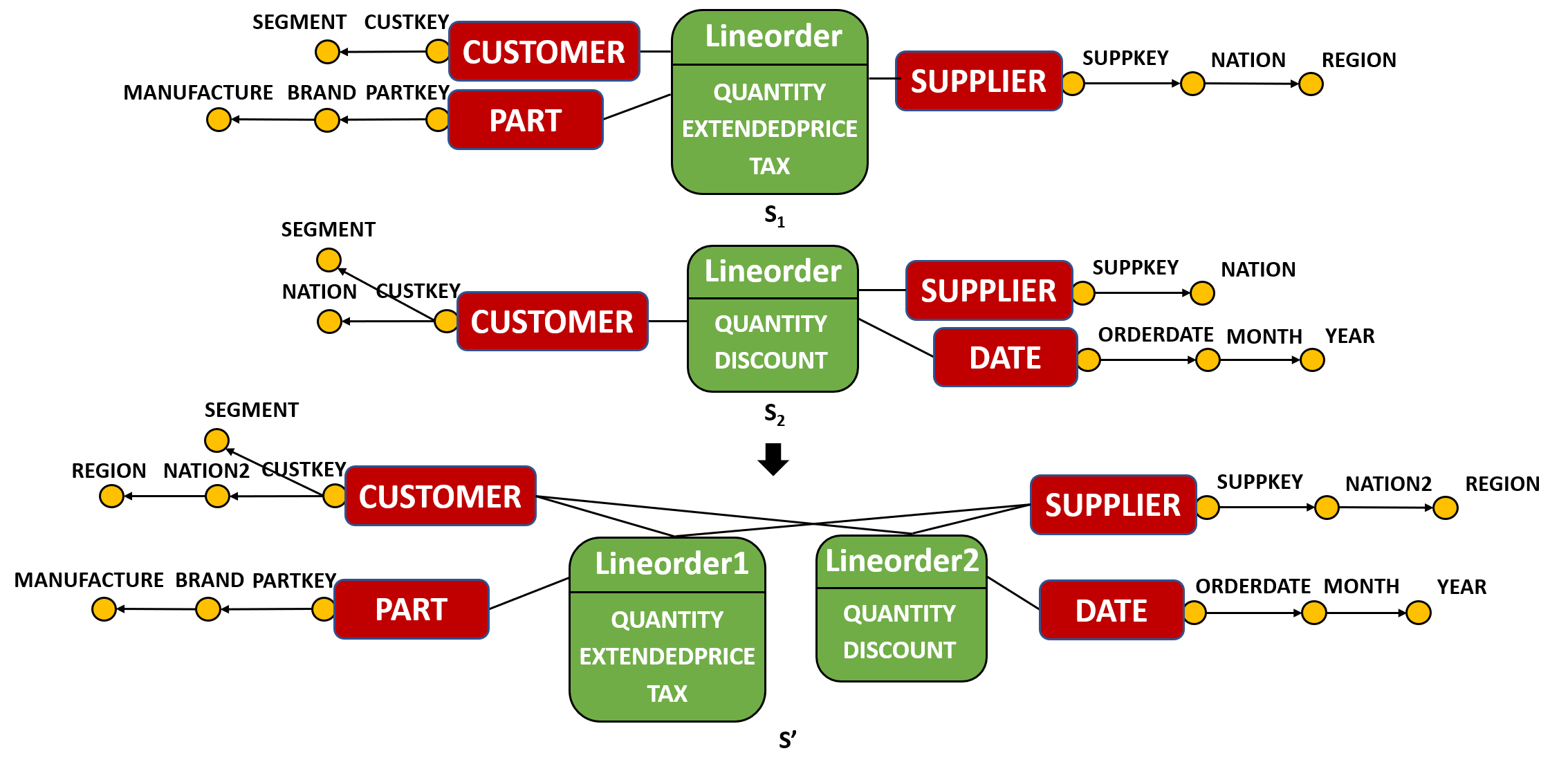}
 \caption{Constellation schema generation} 
 \label{test2}
\end{figure}
At the schema level, the second test generates a constellation schema like shown in Figure \ref{test2}. The run time is 32.13s. As expected, the 2 dimensions $Customer$, $Supplier$ of the original DWs are merged, the other dimension and fact tables are not merged. The dimension $Customer$ gains a new attribute $Region$ by the merging between $S_1.Supplier$ and $S_2.Customer$. We can see that the hierarchy $Custkey \rightarrow nation$ of $Customer$ which should be in the merged schema is deleted because there is no tuple which belongs to this hierarchy but not to $Custkey \rightarrow nation \rightarrow Region$. The hierarchy $Suppkey \rightarrow nation$ of $Supplier$ is removed due to the same reason.

At the instance level, the data of experiment can be found in github. They also meet $N' = N_1 + N_2 - N_{\cap}$. There are empty values of the attribute $Region$ in the dimension $Customer$ and $Supplier$ which are completed which meet $N’ = N_1 + N_2 + N_+$.

We got the results conforming to our expectations in the tests, we can thus conclude that our algorithms work well for the different cases discussed at both schema and instance levels.

\section{Conclusion and future work}
In this paper, we define an automatic approach to merge two different star schema-modeled DWs, by merging multidimensional schema elements including hierarchies, dimensions and facts at the schema and instance levels. We define the corresponding algorithms, which consider different cases. Our algorithms are implemented and illustrated by various examples.

Since we only discuss the merging of DWs modeled as star schemata in this paper, which is only one  (albeit common) possible DW design, we plan to extend our approach by adding the merging of DWs modelled as constellation schemata in the future. There may also be so-called weak attributes in DW components. Thus, we will consider them in future work. Our goal is to provide a complete approach that is integrated in our previous work concerning the automatic integration of tabular data in DWs.
\begin{acks}
The research depicted in this paper is funded by the French National Research Agency (ANR), project ANR-19-CE23-0005 BI4people (Business Intelligence for the people).
\end{acks}
\bibliographystyle{ACM-Reference-Format}
\bibliography{biblio}


\begin{thebibliography}{15}


\ifx \showCODEN    \undefined \def \showCODEN     #1{\unskip}     \fi
\ifx \showDOI      \undefined \def \showDOI       #1{#1}\fi
\ifx \showISBNx    \undefined \def \showISBNx     #1{\unskip}     \fi
\ifx \showISBNxiii \undefined \def \showISBNxiii  #1{\unskip}     \fi
\ifx \showISSN     \undefined \def \showISSN      #1{\unskip}     \fi
\ifx \showLCCN     \undefined \def \showLCCN      #1{\unskip}     \fi
\ifx \shownote     \undefined \def \shownote      #1{#1}          \fi
\ifx \showarticletitle \undefined \def \showarticletitle #1{#1}   \fi
\ifx \showURL      \undefined \def \showURL       {\relax}        \fi
\providecommand\bibfield[2]{#2}
\providecommand\bibinfo[2]{#2}
\providecommand\natexlab[1]{#1}
\providecommand\showeprint[2][]{arXiv:#2}

\bibitem[\protect\citeauthoryear{Ariyan and Bertossi}{Ariyan and
  Bertossi}{2011}]%
        {Sina11}
\bibfield{author}{\bibinfo{person}{Sina Ariyan} {and} \bibinfo{person}{Leopoldo
  Bertossi}.} \bibinfo{year}{2011}\natexlab{}.
\newblock \showarticletitle{Structural Repairs of Multidimensional Databases}.
  In \bibinfo{booktitle}{\emph{Inter. Workshop on Foundations of Data
  Management}}, Vol.~\bibinfo{volume}{748}.
\newblock


\bibitem[\protect\citeauthoryear{Banek, Vrdoljak, Tjoa, and Sko{\v{c}}ir}{Banek
  et~al\mbox{.}}{2007}]%
        {Banek07}
\bibfield{author}{\bibinfo{person}{Marko Banek}, \bibinfo{person}{Boris
  Vrdoljak}, \bibinfo{person}{A.~Min Tjoa}, {and} \bibinfo{person}{Zoran
  Sko{\v{c}}ir}.} \bibinfo{year}{2007}\natexlab{}.
\newblock \showarticletitle{Automating the Schema Matching Process for
  Heterogeneous Data Warehouses}. In \bibinfo{booktitle}{\emph{Data Warehousing
  and Knowledge Discovery}}. \bibinfo{pages}{45--54}.
\newblock


\bibitem[\protect\citeauthoryear{Bergamaschi, Olaru, Sorrentino, and
  Vincini}{Bergamaschi et~al\mbox{.}}{2011}]%
        {Bergamaschi11}
\bibfield{author}{\bibinfo{person}{S. Bergamaschi}, \bibinfo{person}{M. Olaru},
  \bibinfo{person}{S. Sorrentino}, {and} \bibinfo{person}{M. Vincini}.}
  \bibinfo{year}{2011}\natexlab{}.
\newblock \showarticletitle{Semi-automatic Discovery of Mappings Between
  Heterogeneous Data Warehouse Dimensions}.
\newblock \bibinfo{journal}{\emph{J. of Computing and Information Technology}}
  (\bibinfo{date}{dec} \bibinfo{year}{2011}), \bibinfo{pages}{38--46}.
\newblock


\bibitem[\protect\citeauthoryear{Bernstein, Madhavan, and Rahm}{Bernstein
  et~al\mbox{.}}{2011}]%
        {Philip11}
\bibfield{author}{\bibinfo{person}{Philip~A. Bernstein},
  \bibinfo{person}{Jayant Madhavan}, {and} \bibinfo{person}{Erhard Rahm}.}
  \bibinfo{year}{2011}\natexlab{}.
\newblock \showarticletitle{Generic Schema Matching, Ten Years Later}.
\newblock \bibinfo{journal}{\emph{Proc. VLDB Endow.}} \bibinfo{volume}{4},
  \bibinfo{number}{11} (\bibinfo{date}{aug} \bibinfo{year}{2011}),
  \bibinfo{pages}{695–701}.
\newblock


\bibitem[\protect\citeauthoryear{Elamin, Alzaidi, and Feki}{Elamin
  et~al\mbox{.}}{2018}]%
        {Elamin18}
\bibfield{author}{\bibinfo{person}{Elhaj Elamin}, \bibinfo{person}{Amer
  Alzaidi}, {and} \bibinfo{person}{Jamel Feki}.}
  \bibinfo{year}{2018}\natexlab{}.
\newblock \showarticletitle{A Semantic Resource Based Approach for Star Schemas
  Matching}.
\newblock \bibinfo{journal}{\emph{IJDMS}} \bibinfo{volume}{10},
  \bibinfo{number}{6} (\bibinfo{date}{dec} \bibinfo{year}{2018}).
\newblock


\bibitem[\protect\citeauthoryear{Elavarasi, Akilandeswari, and
  Menaga}{Elavarasi et~al\mbox{.}}{2014}]%
        {Elavarasi14}
\bibfield{author}{\bibinfo{person}{S.~Anitha Elavarasi}, \bibinfo{person}{J.
  Akilandeswari}, {and} \bibinfo{person}{K. Menaga}.}
  \bibinfo{year}{2014}\natexlab{}.
\newblock \showarticletitle{A Survey on Semantic Similarity Measure}.
\newblock \bibinfo{journal}{\emph{Inter. J. of Research in Advent Technology}}
  \bibinfo{volume}{2} (\bibinfo{date}{mar} \bibinfo{year}{2014}).
\newblock


\bibitem[\protect\citeauthoryear{Feki, Majdoubi, and Gargouri}{Feki
  et~al\mbox{.}}{2005}]%
        {Jamel05}
\bibfield{author}{\bibinfo{person}{Jamel Feki}, \bibinfo{person}{Jihen
  Majdoubi}, {and} \bibinfo{person}{Faïez Gargouri}.}
  \bibinfo{year}{2005}\natexlab{}.
\newblock \showarticletitle{A Two-Phase Approach for Multidimensional Schemes
  Integration}. In \bibinfo{booktitle}{\emph{17th Inter. Conference on Software
  Engineering and Knowledge Engineering}}. \bibinfo{pages}{498--503}.
\newblock


\bibitem[\protect\citeauthoryear{Kwakye, Kiringa, and Viktor}{Kwakye
  et~al\mbox{.}}{2013}]%
        {Kwakye13}
\bibfield{author}{\bibinfo{person}{M. Kwakye}, \bibinfo{person}{I. Kiringa},
  {and} \bibinfo{person}{H.~L. Viktor}.} \bibinfo{year}{2013}\natexlab{}.
\newblock \showarticletitle{Merging Multidimensional Data Models: A Practical
  Approach for Schema and Data Instances}. In \bibinfo{booktitle}{\emph{5th
  Inter. Conference on Advances in Databases, Data, and Knowledge
  Applications}}.
\newblock


\bibitem[\protect\citeauthoryear{March and Hevner}{March and Hevner}{2007}]%
        {Salvatore07}
\bibfield{author}{\bibinfo{person}{Salvatore~T. March} {and}
  \bibinfo{person}{Alan~R. Hevner}.} \bibinfo{year}{2007}\natexlab{}.
\newblock \showarticletitle{Integrated decision support systems: A data
  warehousing perspective}.
\newblock \bibinfo{journal}{\emph{Decis. Support Syst.}} \bibinfo{volume}{43},
  \bibinfo{number}{3} (\bibinfo{date}{apr} \bibinfo{year}{2007}),
  \bibinfo{pages}{1031 -- 1043}.
\newblock


\bibitem[\protect\citeauthoryear{Meng, Huang, and Gu}{Meng
  et~al\mbox{.}}{2013}]%
        {Meng13}
\bibfield{author}{\bibinfo{person}{Lingling Meng}, \bibinfo{person}{Runqing
  Huang}, {and} \bibinfo{person}{Junzhong Gu}.}
  \bibinfo{year}{2013}\natexlab{}.
\newblock \showarticletitle{A review of semantic similarity measures in
  wordnet}.
\newblock \bibinfo{journal}{\emph{IJHIT}}  \bibinfo{volume}{6}
  (\bibinfo{date}{jan} \bibinfo{year}{2013}).
\newblock


\bibitem[\protect\citeauthoryear{Olaru and Vincini}{Olaru and Vincini}{2012}]%
        {Olaru12}
\bibfield{author}{\bibinfo{person}{Marius-Octavian Olaru} {and}
  \bibinfo{person}{Maurizio Vincini}.} \bibinfo{year}{2012}\natexlab{}.
\newblock \showarticletitle{A Dimension Integration Method for a Heterogeneous
  Data Warehouse Environment}. In \bibinfo{booktitle}{\emph{Inter. Conf. on
  Data Communication Networking, e-Business and Optical Communication
  Systems}}.
\newblock


\bibitem[\protect\citeauthoryear{Quix, Kensche, and Li}{Quix
  et~al\mbox{.}}{2007}]%
        {Quix07}
\bibfield{author}{\bibinfo{person}{Christoph Quix}, \bibinfo{person}{David
  Kensche}, {and} \bibinfo{person}{Xiang Li}.} \bibinfo{year}{2007}\natexlab{}.
\newblock \showarticletitle{Generic Schema Merging}. In
  \bibinfo{booktitle}{\emph{Advanced Information Systems Engineering}}.
  \bibinfo{pages}{127--141}.
\newblock


\bibitem[\protect\citeauthoryear{Ravat, Teste, Tournier, and Zurfluh}{Ravat
  et~al\mbox{.}}{2008}]%
        {Ravat08}
\bibfield{author}{\bibinfo{person}{Franck Ravat}, \bibinfo{person}{Olivier
  Teste}, \bibinfo{person}{Ronan Tournier}, {and} \bibinfo{person}{Gilles
  Zurfluh}.} \bibinfo{year}{2008}\natexlab{}.
\newblock \showarticletitle{Algebraic and Graphic Languages for OLAP
  Manipulations}.
\newblock \bibinfo{journal}{\emph{Inter. J. of Data Warehousing and Mining}}
  \bibinfo{volume}{4} (\bibinfo{date}{jan} \bibinfo{year}{2008}),
  \bibinfo{pages}{17--46}.
\newblock


\bibitem[\protect\citeauthoryear{Romero and Abelló}{Romero and
  Abelló}{2009}]%
        {Romero09}
\bibfield{author}{\bibinfo{person}{Oscar Romero} {and} \bibinfo{person}{Alberto
  Abelló}.} \bibinfo{year}{2009}\natexlab{}.
\newblock \showarticletitle{A Survey of Multidimensional Modeling
  Methodologies}.
\newblock \bibinfo{journal}{\emph{Inter. J. of Data Warehousing and Mining}}
  \bibinfo{volume}{5}, \bibinfo{number}{2} (\bibinfo{date}{apr}
  \bibinfo{year}{2009}).
\newblock


\bibitem[\protect\citeauthoryear{Torlone}{Torlone}{2008}]%
        {Torlone08}
\bibfield{author}{\bibinfo{person}{Riccardo Torlone}.}
  \bibinfo{year}{2008}\natexlab{}.
\newblock \showarticletitle{Two approaches to the integration of heterogeneous
  data warehouses}.
\newblock \bibinfo{journal}{\emph{Distributed and Parallel Databases}}
  \bibinfo{volume}{23} (\bibinfo{date}{feb} \bibinfo{year}{2008}),
  \bibinfo{pages}{69–97}.
\newblock


\end{thebibliography}

\end{document}